\documentclass[%
 aip,
 amsmath,amssymb,
 reprint,%
]{revtex4-1}

\usepackage{graphicx}% Include figure files
\usepackage{dcolumn}% Align table columns on decimal point
\usepackage{bm}% bold math
\usepackage[utf8]{inputenc}
\usepackage[T1]{fontenc}
\usepackage{mathptmx}
\usepackage[breaklinks=true,colorlinks=true,linkcolor=blue,urlcolor=blue,citecolor=blue]{hyperref}% add hypertext capabilities

\usepackage{xurl}      % how to make links have hyperlinks?
\usepackage{makeidx}
\usepackage{empheq}  % multi-line boxes
\usepackage{color}
\usepackage{subcaption}    
\usepackage{booktabs} % for nicer rules in the table  (added March 4, 2025)
\usepackage{placeins}  % added Nov 13 '25 to allow column-spanning objects to sit at the bottom of a page
\usepackage[table]{xcolor}

\usepackage{color} % used for the \color{} command
\usepackage{soul}

\graphicspath{{Images/}}

\begin{document}

\preprint{AIP/123-QED}

\title[Revisiting confinement scalings and fusion performance with a perspective optimized for extrapolation]{Revisiting confinement scalings and fusion performance with a perspective optimized for extrapolation}

\author{Jalal Butt}
\affiliation{Princeton University}
\author{Geert Verdoolaege}
\affiliation{Ghent University}
\author{Stanley M. Kaye}
\affiliation{Princeton Plasma Physics Laboratory}
\author{Egemen Kolemen}
\email[Corresponding author: ]{ekolemen@princeton.edu}
\affiliation{Princeton University}
\affiliation{Princeton Plasma Physics Laboratory}

\date{20 April 2026}

\begin{abstract}

\noindent Recent advances in high-temperature-superconductor technology have made substantially higher toroidal magnetic fields technologically accessible, reopening the design space for compact, high-field tokamak reactors. Because reactor performance projections remain anchored to empirical confinement scalings, the recent update to the ITPA global H-mode confinement database raises an important question: what does the present experimental record and its uncertainty imply for the path to reactor-grade fusion performance? In this work, we revisit confinement extrapolation from an explicitly extrapolation-oriented perspective and, to complement its implications in terms of a direct reactor performance measure, present a cross-machine empirical scaling for fusion power. We systematically search for a minimally complex confinement scaling that optimizes the tradeoff between variance capture and extrapolative robustness. We find that low-order models centered near $N=3$ to $N=4$ optimize this tradeoff, with plasma current, machine size, heating power, and elongation emerging as the dominant engineering levers, together with an empirically inferred confinement penalty associated with metallic walls. Recast in reactor-performance terms, the results indicate that both the fusion triple product and fusion power are governed primarily by plasma current: the triple product scales approximately as $I_p^2$, and the empirical fusion power scaling exhibits a similarly near-quadratic dependence over a survey of the highest performing discharges across several machines. Projecting to reactors, these results suggest that high-field devices with metal walls may require higher plasma current than standard IPB98$(y,2)$-based expectations imply, and that gigawatt-class tokamak performance likely demands operation at $I_p \gtrsim 20\mathrm{MA}$.
\end{abstract}

\maketitle

\section{Introduction}

\begin{figure}
    \centering
    \includegraphics[width=0.9\linewidth]{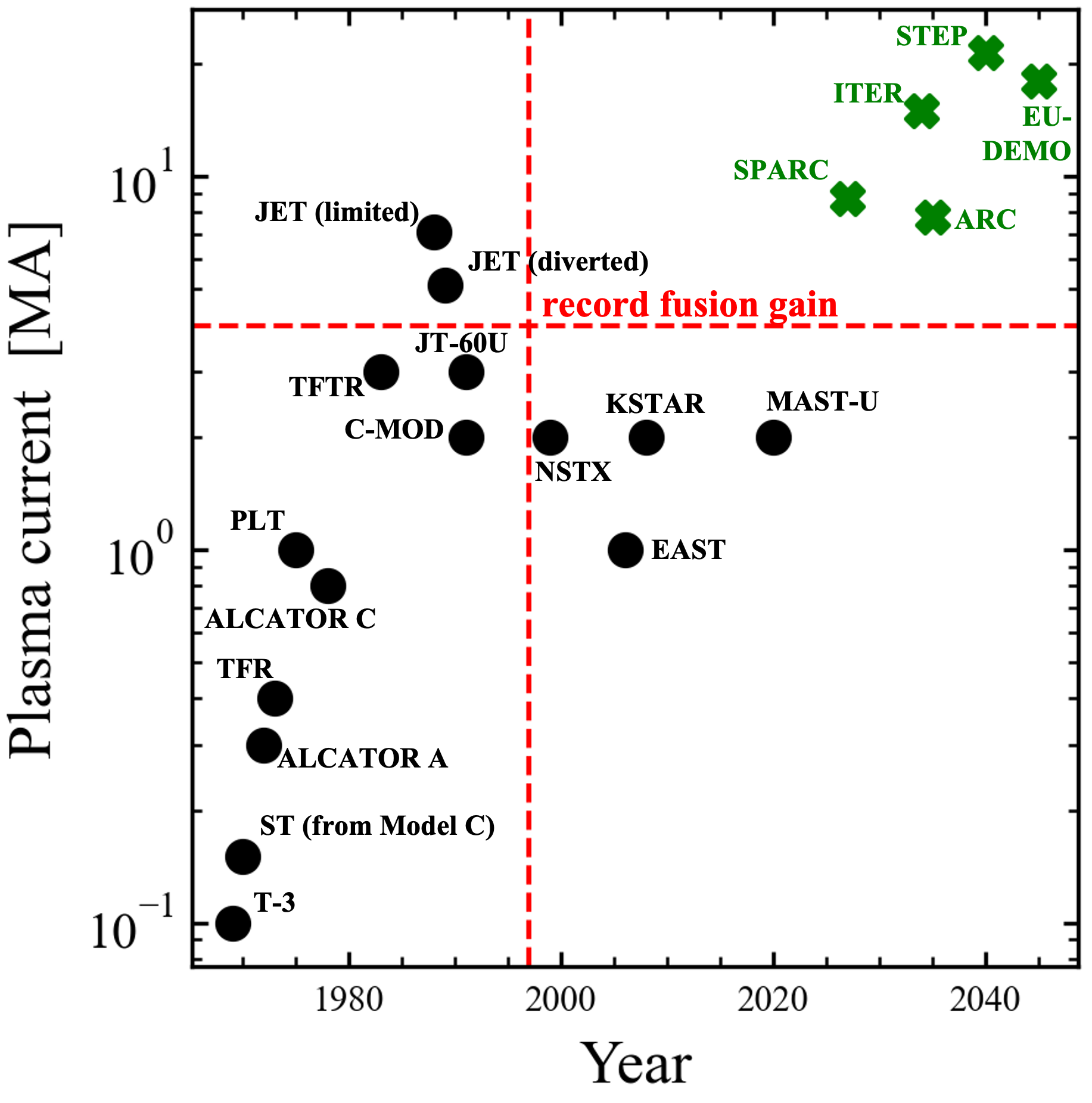}
    \captionsetup{width=\linewidth, font= footnotesize}
    \caption{Peak plasma current achieved in major machine since the T-3 tokamak in 1969 to the US reactor concept ARC self-reported to come online in the mid 2030s and EU-DEMO in the late 2040s. Note that JET's limiter operation up to 7.1MA was merely an operational boundary demonstration that was not explored \cite{bertoliniEngineeringExperienceJET1997}.}
    \label{fig: Ip vs time}
\end{figure}

\noindent Through the 1990s, records in fusion performance were regularly improved upon\cite{wurzel_continuing_2025, wurzel_progress_2022}, culminating in the JET divertor upgrade that enabled operation up to $I_p=5.1$MA \cite{keilhackerFusionPhysicsProgress1999, bertoliniEngineeringExperienceJET1997}. Within a few years, the magnetic confinement fusion (MCF) gain reached its present record of $Q=0.67$ \cite{keilhacker_high_1999}. As will be argued in this letter, progress in fusion performance requires, to first-order, access to higher plasma current. In this light, the decades-long stagnation in tokamak performance can be viewed as a stagnation in achievable $I_p$, illustrated in figure \ref{fig: Ip vs time}. 

One way to quantify fusion performance is through the ignition parameter\cite{lawsonCriteriaPowerProducing1957} $\langle nT\rangle \tau$, which measures whether fueling nuclei remain confined in high-probability-of-reaction conditions—high $\langle nT\rangle$ sustained for sufficiently long $\tau$. While density and temperature can often be constrained by operational limits and design choices, a reliable, first-principles prediction of the energy confinement time $\tau_E$ for reactor-scale plasmas remains a central and unresolved challenge.

In the absence of a widely accepted first-principles theory of global transport, empirical confinement scalings have long been an indispensable tool for projecting tokamak performance \cite{goldston_energy_1984, yushmanov_scalings_1990, iter_physics_expert_groups_on_confinement_and_transport_and_confinement_modelling_and_database_plasma_1999}. These scalings are typically judged by their ability to reproduce systematic trends in existing confinement data, and by that measure they often appear highly successful. However, recent analysis of the updated global H-mode confinement database \cite{verdoolaege_updated_2021} shows that several scaling exponents carry substantial in-distribution uncertainty. When propagated to reactor regimes, this uncertainty produces a significantly broader range of projected performance than earlier, less robust methodologies made apparent. This underscores the need to reassess how confinement scalings are constructed. In a related vein, we also pursue empirical scalings for fusion performance.

In the work of deriving empirical confinement scalings, increasing model complexity generally improves in-sample fit, but only up to the point where additional parameters begin to capture statistical noise rather than real physical variation. The resulting choice of model order is therefore a bias-variance tradeoff: overly simple models risk systematic bias, whereas overly complex models can exhibit high variance under extrapolation. Because different prior assumptions can lead to somewhat different fitted scalings, the choices made ought to be stated explicitly and justified quantitatively rather than implicitly inherited from historical practice. In this work, we adopt a methodology that we believe best simulates the present extrapolation problem to identify the optimal model adequate for projection within the scope of the task. We attempt our honest best and present our findings as an insightful guide rather than explicit predictions.

\section{Fitting an optimally low order confinement time scaling}
\label{section: Fitting an optimally low order confinement scaling}

\begin{figure}
    \centering
    \includegraphics[width=0.9\linewidth]{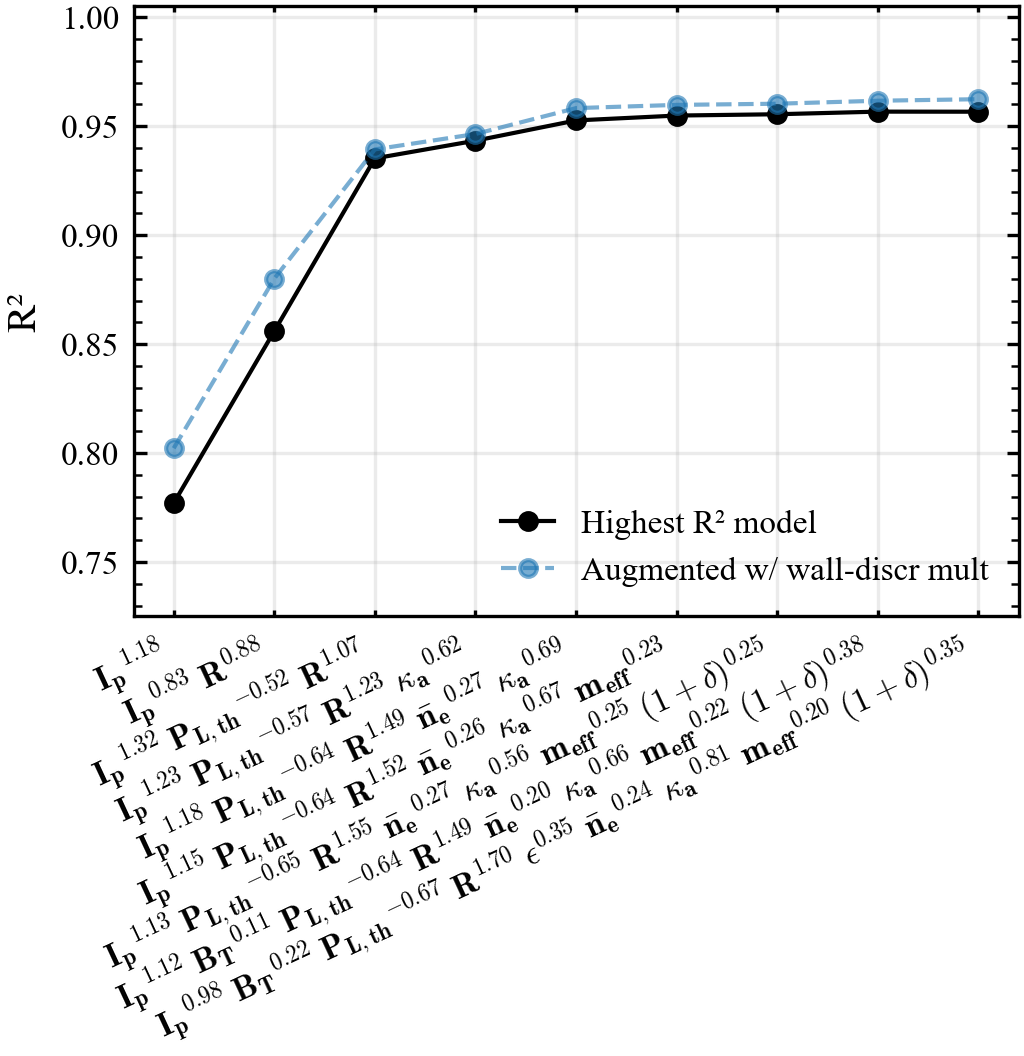}
    \captionsetup{width=\linewidth, font= footnotesize}
    \caption{Confinement-time scalings with the highest $R^2$ among all parameter combinations in the ITPA H-mode database for model orders $N=1$ to $N=9$. The solid curve corresponds to the fits obtained without wall discrimination. Also shown, for consistent visual comparison, are the same $N$-parameter scalings retroactively augmented by a wall-discriminating multiplicative coefficient obtained by fitting only that coefficient while keeping the rest of the model fixed. The natively fit wall-discriminating scalings are presented in appendix sections \ref{appendix: Full $N$-parameter models' details} and \ref{appendix:  Wall-discriminating coefficient variability with model complexity}.}
    \label{fig: cascade tauE}
\end{figure}

\noindent To address these extrapolative concerns, we seek a minimally complex thermal energy confinement scaling that captures most of the database variance while approaching the performance of the traditional higher-order parameter sets.

Empirical confinement scalings summarize multi-machine data with a power-law relation in a conventional set of engineering variables. Their utility lies less in assigning unique physical meaning to any particular fitted exponent than in providing a compact empirical representation of confinement trends for comparison, projection, and reactor scoping. Although adding parameters generally improves in-sample fit, it need not improve -- and may degrade -- extrapolative robustness. The usual engineering parameter sets therefore reflect a mixture of empirical adequacy, physical interpretability, and historical convention. Motivated by renewed reactor design interest and the updated database, scaling, and uncertainty analysis, we revisit this truncation systematically and note the spread in predictions associated with model-order choice.

We search for a model that captures the database variance with minimal complexity by considering every combination of engineering parameters available in the standard set of the DB5.2.3 database \cite{verdoolaege_updated_2021}, beginning with single-parameter fits and extending up to the usual nine to ten parameters. The highest $R^2$ values obtained from weighted log-log (power-law) least-squares fits for each model order $N$ are shown in figure \ref{fig: cascade tauE}. As expected, fit quality improves monotonically as parameters are added, but the gain in variance capture drops sharply beyond about $N=3$, indicating a knee in the tradeoff between in-database fit quality and the extrapolative robustness of higher-complexity models.

To identify this truncation point more systematically, we carried out the analysis documented in appendix section \ref{appendix: Systematic determination of $N-$parameter model noninferiority to highest complexity model}. There, model forms of increasing order are trained only on lower-$\tau_E$ data (0th-30th percentile range, up to 0th-60th percentile range) and evaluated on their ability to predict the highest-$\tau_E$ portion of the database (85th-95th percentile range). Relative to the highest-complexity reference model ($N=9$), this percentile-based prior analysis identifies the $N=3$ model as the lowest-order model that is not meaningfully inferior for prediction in the 85-95th percentile range.

However, prediction of high-$\tau_E$ in an inter-machine database is still not the extrapolation task we believe to be of ultimate interest. The intended application is prediction of a reactor-scale device that has not yet been built. For that reason, we performed a second, complementary extrapolation analysis in which the prior is constructed in a way that more closely mirrors what we interpret as the most relevant question at hand. Rather than asking how well a model trained on lower-performance discharges predicts higher-performance discharges, we ask how well models of varying complexity would have predicted the next tokamak in the historical development of the field using only information from its predecessors. We believe this machine-based prior construction is a better aligned test of model complexity for reactor extrapolation than within-inter-machine-database alone. To implement this test, we chronologically order the machines in the database by start-date, and, for each step in the sequence, use data from all preceding machines to predict the next one. This analysis finds that extrapolative RMSE improves to a minimum at $N=3$ and then worsens as further parameters are added, while the corresponding bias metric reaches its minimum at $N=5$. Thus, although the precise optimum depends modestly on the metric, both extrapolation frameworks independently place the preferred model complexity in the $N=3$ to $N=4$ low-order range.

Taken together, the extrapolation tasks of predicting higher-$\tau_E$ regimes from lower-$\tau_E$ data, predicting future machines from prior machines, and alternative non-chronological machine hierarchy tests (e.g. performance hierarchy) documented in the appendix point us to essentially the same conclusion: the most appropriate model complexity lies near $N=3$ to $N=4$. We therefore adopt the $N=3$ model as the principal model of choice, while treating the marginally more complex $N=4$ model as a secondary reference check. In this sense, these optimally low-order models provide a task-motivated sanity check on the principal confinement-time model employed in the reactor design.

The optimal low-order (3-parameter) scaling is,

\begin{align}
    \tau_{E,th} =0.123\;  I_p^{1.31} \, P_{l,th}^{-0.49} \, R_{geo}^{1.05} \; \times 0.896^{\text{low-Z: 0;  high-Z:  1}},
    \label{eq: 3-param tauE}
\end{align}

where the wall-discriminating indicator function yields a coefficient of unity for low-Z walls and 0.896 for high-Z walls, effectively acting as a confinement penalty of approximately 10\% for high-Z walls. Further analysis of the wall-discriminating coefficient is supplied in the appendix, section \ref{appendix:  Wall-discriminating coefficient variability with model complexity}, from which we note that the penalty converges to approximately 15\% for models $N>5$ parameters. We also note that the $N=3$ model has the lowest RBAYES \cite{verdoolaege_updated_2021} error of models with neighboring complexity at least up to $N=5$, presented in table \ref{table: incremental_scalings_aug_uncertainties}. 

This optimally low order model reproduces some familiar insight: the strongest levers on confinement we are most confident in are the input power, plasma current, and geometric major radius. These variables also appear as three of four of the largest levers of confinement in IPB98(y,2)\cite{iterphysicsexpertgrouponconfinementandtransportChapter2Plasma1999} and ITPA20\cite{verdoolaege_updated_2021}, where the top three are plasma current, major radius, and elongation. The $N=4$ model states the elongation to be the fourth largest lever. However, we recall that the advanced bayesian techniques employed by Verdoolaege et al find elongation to exhibit substantial uncertainty, both in the ITPA20 analysis paper \cite{verdoolaege_updated_2021} and per table \ref{table: incremental_scalings_aug_uncertainties}.

\begin{table*}[ht]
\centering
\begin{tabular}{lccccc}
\hline
Machine  & $P_{\text{fus}}$ [MW] & $I_p$ [MA] & $B_T$ [T] & $A$ & $H\; [\text{MW}/\sqrt{\text{MA}}]$ \\
\hline
JET DTE1 (C wall; 1997) \cite{keilhacker_high_1999}              & 16.1  & 4.1 & 3.6  & 3.0 & 0.7 \\
JET DTE2 (ILW; 2021) \cite{hobirk_jet_2023}                     & 10    & 2.5 & 3.45 & 3.0 & 1.6 \\
TFTR (1994) \cite{hawryluk_fusion_1998}                         & 10.7  & 2.7 & 5.6  & 2.9 & 1.2 \\
JT-60U (DT-eq, H-mode; 1997) \cite{kamada_long_1999}            & 2.4   & 1.5 & 3.6  & 4.1 & 1.0 \\
JT-60U (DT-eq, RS; 1996) \cite{ishidaAchievementHighFusion1997} & 10.7  & 2.79 & 4.34 & 4.4 & 1.1 \\
SPARC H-mode (under construction) \cite{creely_overview_2020}   & 140   & 8.7 & 12.2 & 3.2 & 1.1 \\
ITER Baseline (under construction) \cite{iter_physics_expert_groups_on_confinement_and_transport_and_confinement_modelling_and_database_plasma_1999} & 500 & 15.0 & 5.3 & 3.1 & 1.1 \\
DIII-D (DT-eq, SH; 2019)\cite{snyder_high_2019}                 & 4.80  & 1.95 & 2.2 & 2.8 & 1.1 \\
DIII-D (DT-eq, NCS; 1997) \cite{lazarus_higher_1997}            & 3.3   & 2.20 & 2.5 & 2.8 & 0.6 \\
\hline
\end{tabular}
\captionsetup{width=\linewidth, font=footnotesize}
\caption{Comparison of D-T equivalent fusion power, self-reported projection of fusion power for designs under mature stage construction, or achieved fusion powers (all designated $P_{fus}$), along with the discharge's $I_p$, aspect ratio $A$, and coefficient on $H$-factor (ratio of produced or expected fusion power to fusion power predicted by the full dataset's best-fitting, single-variable power-law scaling as presented in figure \ref{fig: Pfus scaling}).}
\label{table:  Pfus}
\end{table*}

What insight can we draw from this optimal low-order confinement time scaling in terms of fusion performance? A convenient form within approximate range of our optimal empirical fit is $\tau_E= H\;c_1(I_p R^{3/2})/P_{L,th}^{1/2}$. This allows simplification to find an explicit expression of the triple product. Assuming a circular poloidal cross-section,
\begin{align*}
    \tau_E &= 0.123\,H  \frac{I_pR^{3/2}}{(nTV/\tau_E)^{1/2}} \\
    &= \frac{0.123\,H  }{(2 \pi^2)^{1/2}} \frac{I_pR^{3/2}}{(nT(Ra^2)/\tau_E)^{1/2}},
\end{align*}
which gives,
\begin{align}
    nT\tau_E = c \; H^2\, I_p^{2}\, A^2.
    \label{eq: H^2 * Ip^2 * A^2}
\end{align}

where $I_p$ is in MA, $nT\tau_E$ is in $keV\, m^{-3} \, s$, and $c$ is a coefficient. Remarkably, this empirical result, culminating decades of H-mode research, nearly reproduces the 42-year-old fusion triple product scaling for L-mode first derived by Goldston \cite{goldston_energy_1984}, 

\begin{align*}
    nT\tau_E = 1.4 \times 10^{17} \; H^2\, I_p^2\, A^{2.5} \, a^{-0.24},
\end{align*}

where $I_p$ is in MA and $nT\tau_E$ is in $keV\, m^{-3} \, s$. The same argument was made by Wesson in an H-mode retrospective\cite{wesson_science_1999}. The same scaling for $nT\tau_E$ is recovered for the $N=4$ case by accounting for plasma elongation when approximating $\alpha_\kappa$ as $0.5$ and applying an elongation correction to the cross-section. 

Just as the L-mode result from 42 years ago indicated, its high-confinement sibling \textit{H-mode} lucidly states that the triple product measure of fusion performance scales quadratically with plasma current, aspect ratio, and H-factor enhancement to optimal low order. While this work considers all H-mode data in the latest ITPA H-mode standard dataset, we focus our conclusions on conventional aspect-ratio tokamaks, since spherical tokamaks occupy a low-aspect-ratio regime whose empirical confinement dependencies appear distinct from those of conventional tokamaks \cite{kayeThermalConfinementTransport2021}. 
The optimal order scaling analysis conducted herein shows that the fusion triple product and fusion power scale with $I_p^2$. Our simple treatment on discriminating the effect of metal walls on confinement suggests that their integrated effect is to decrease the effective H-factor. Considering that our analysis finds $I_p$ to be the single engineering lever for a fixed vessel size on $n T \tau_E$,  this implies that metal wall reactors ought to design for higher $I_p$ to achieve the same $Q$ as one designed with low-Z wall confinement projections.

We contextualize this finding with the recent update to the ITPA global H-mode confinement database. The updated H-mode confinement scaling \cite{verdoolaege_updated_2021} shares many similarities with that of IPB98(y,2), yielding similar baseline predictions for ITER. Using advanced bayesian techniques to extract more robust uncertainties, Verdoolaege et al found that some are quite substantial, up to 100\% for variables including $B_T$ and $(1+\delta)$, and exceeding 200\% for $\epsilon$, and nearly 50\% for $\kappa_a$. When the database is partitioned by wall material, two different branches emerge. The nonmetal-wall subset largely reproduces the nominal IPB98(y,2) and ITPA20 predictions for ITER and SPARC, whereas the metal-wall subset predicts distinctly lower confinement times than either scaling, though likely with greater machine bias given the smaller number of contributing devices. Given the narrow size range of devices contributing to the \textit{all high-Z} scaling, its weak size dependence may not extrapolate reliably beyond the dataset. For reference, we nevertheless apply this scaling to SPARC, noting its compact size between AUG and JET and the contextual value of a multi-machine metal-wall-only comparison. When the optimal low-order scalings with wall-discrimination are applied, the resulting confinement projections shift away from the full-database and \textit{all low-Z} projections and toward the lower confinement \textit{all high-Z} branch. This suggests that the inferred wall penalty is a nontrivial effect in reactor-relevant extrapolation.

\section{Empirical scaling for fusion power ($P_{fus}$):  establishing an empirical yard-stick for progress in fusion power production}
\label{Section:  Empirical scaling for fusion power ($P_{fus}$):  establishing an empirical yard-stick for progress in fusion power}

The fusion triple product is a convenient measure of performance toward burning conditions, but it is not the most directly relevant metric for fusion performance. The more direct quantity we seek is the fusion power output, for which $nT\tau_E$ serves as an order-scale proxy. So, let us examine the parametric dependence of fusion performance in terms of the more direct measure of fusion power production. A range of expectations of the parametric dependence is offered in appendix section \ref{appendix: Some expectations of fusion power's dependence on standard engineering parameters}, which we briefly summarize here for the reader's convenience. Invoking the $N=3$ optimal low order scaling leads us to expect a fusion power scaling like $H^2 \,I_p^{2}A^2 \frac{P_{aux}}{f_{aux}}$, while another argument leads us to $f_{BS}^2I_p^{4}$. An argument similar to the latter yields $(\kappa\beta_T)^2 B_T^4$. These scaling arguments are not used to preselect variables or otherwise constrain the regression search. 

\begin{figure}
    \centering
    \includegraphics[width=0.9\linewidth]{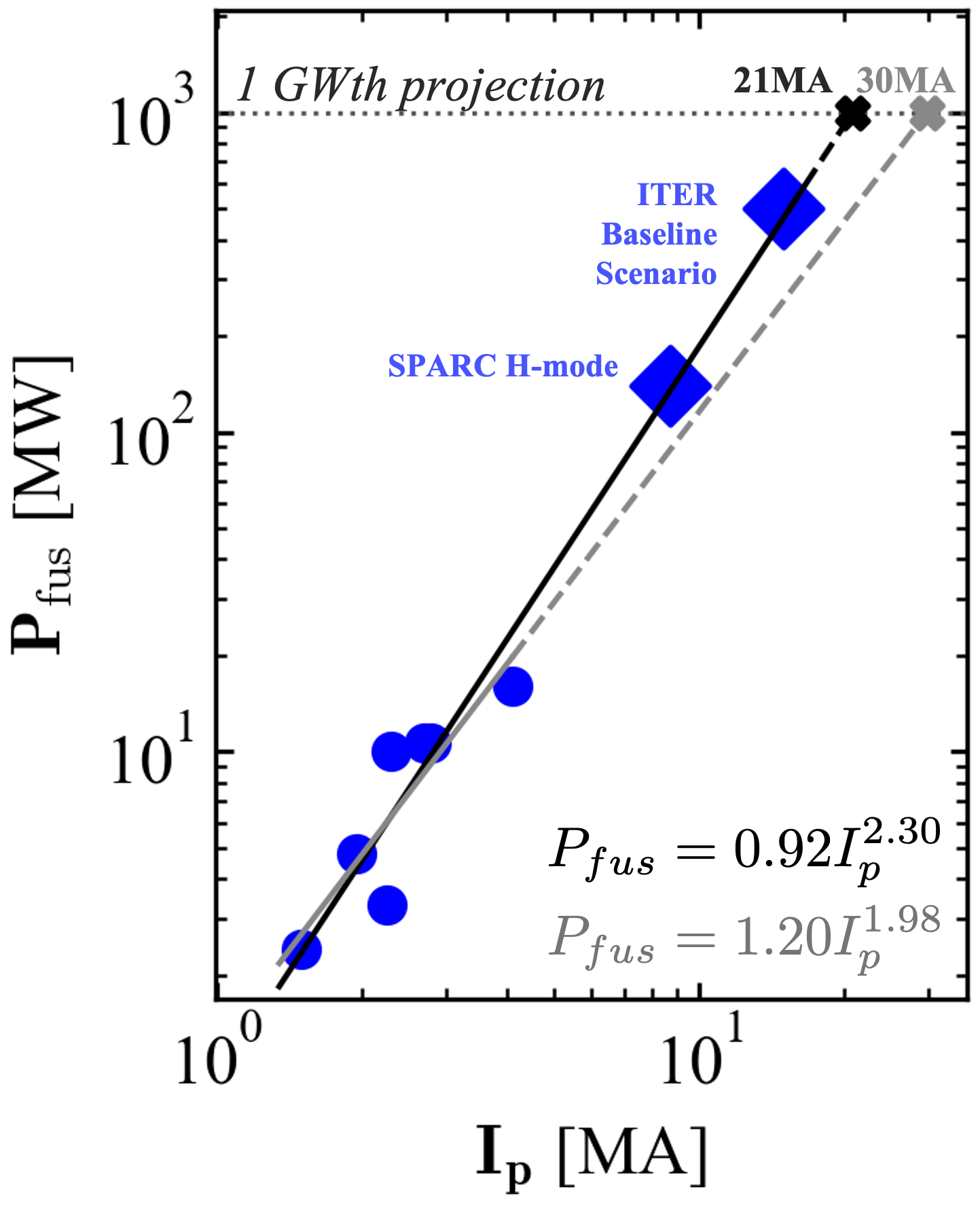}
    \captionsetup{width=\linewidth, font= footnotesize}
    \caption{Fusion power scaling with and without the machines currently under construction (produced from table \ref{table:  Pfus}). The gray curve was fit only to experimentally demonstrated fusion power production, while the black curve also considers designed for machines under construction (ITER Baseline Scenario and SPARC H-mode).}
    \label{fig: Pfus scaling}
\end{figure}

Table \ref{table:  Pfus} is a bulletin of produced and planned fusion powers from a survey of published works. Because this fusion power dataset is much smaller than the confinement database, we do not repeat the same $N$-parameter model-selection exercise here and instead restrict attention to single-variable power-law regressions. Single-variable power-law fits relating $P_{fus}$ to individual standard engineering variables commonly used in confinement scalings show that fusion power is most strongly organized by plasma current, scaling approximately quadratically in $I_p$ with remarkably high goodness of fit ($P_{fus} = 0.92\,I_p^{2.30}$ fits the full survey with log-space $R^2= 0.97$; using only already produced results, the scaling is $P_{fus} = 1.20\,I_p^{1.98}$). This best single-variable fit to fusion power is presented in figure \ref{fig: Pfus scaling}. The variance captured by other standard engineering predictor variables is substantially lower, including for $B_T$, $R_{geo}$, $a_{min}$, $\bar{n}_e$, and $\kappa_a$ (shown in table \ref{table: single_param_fits}). These observations hold both with and without consideration of the planned devices.

It is notable that the variables showing the strongest correlations with fusion power, and the largest fitted exponents, are the same engineering levers that dominate empirical confinement-time scalings, especially plasma current and geometric size. In that context, the absence of $B_T$ as a primary correlate of $P_{fus}$ is not necessarily surprising. If one adopts the simple power-balance argument that $nT$ scales with $\tau_E$, then a weak dependence of $P_{fus}$ on $B_T$ follows from the vanishingly small $B_T$ dependence in standard confinement scalings, including but not limited to the optimal low order scalings presented. This fusion power scaling is therefore broadly consistent with empirical confinement scalings in identifying little direct role for $B_T$ for conventional aspect ratio tokamaks.  Fusion power scaling with $I_p^2$ implies that the fusion gain $Q$ scales with $I_p^2$ at fixed input power.

This regression indicates that $\gtrsim$20MA is needed to achieve 1 GW of thermal power. The fusion power data considered here are taken from the highest performing discharges of each machine, so it has already filtered out under-optimized scenarios and retained only each device’s most successful attempts. As a result, the inferred requirement of roughly $20 \,\mathrm{MA}$ for a gigawatt-class powerplant is not a conservative one. It is more aptly interpreted as an optimistic, empirically-grounded estimate, since alternative strategies for compensating lower $I_p$ have already been implicitly explored in the construction of these data points. In fact, at a minimum, the coefficient on the scaling will be lower if only the thermonuclear component of fusion power is of interest.

\section{Implications for reactor design}
\begin{figure}[ht!]
    \begin{subfigure}[t]{0.99 \linewidth}
        \centering
        \includegraphics[width= 0.975\linewidth]{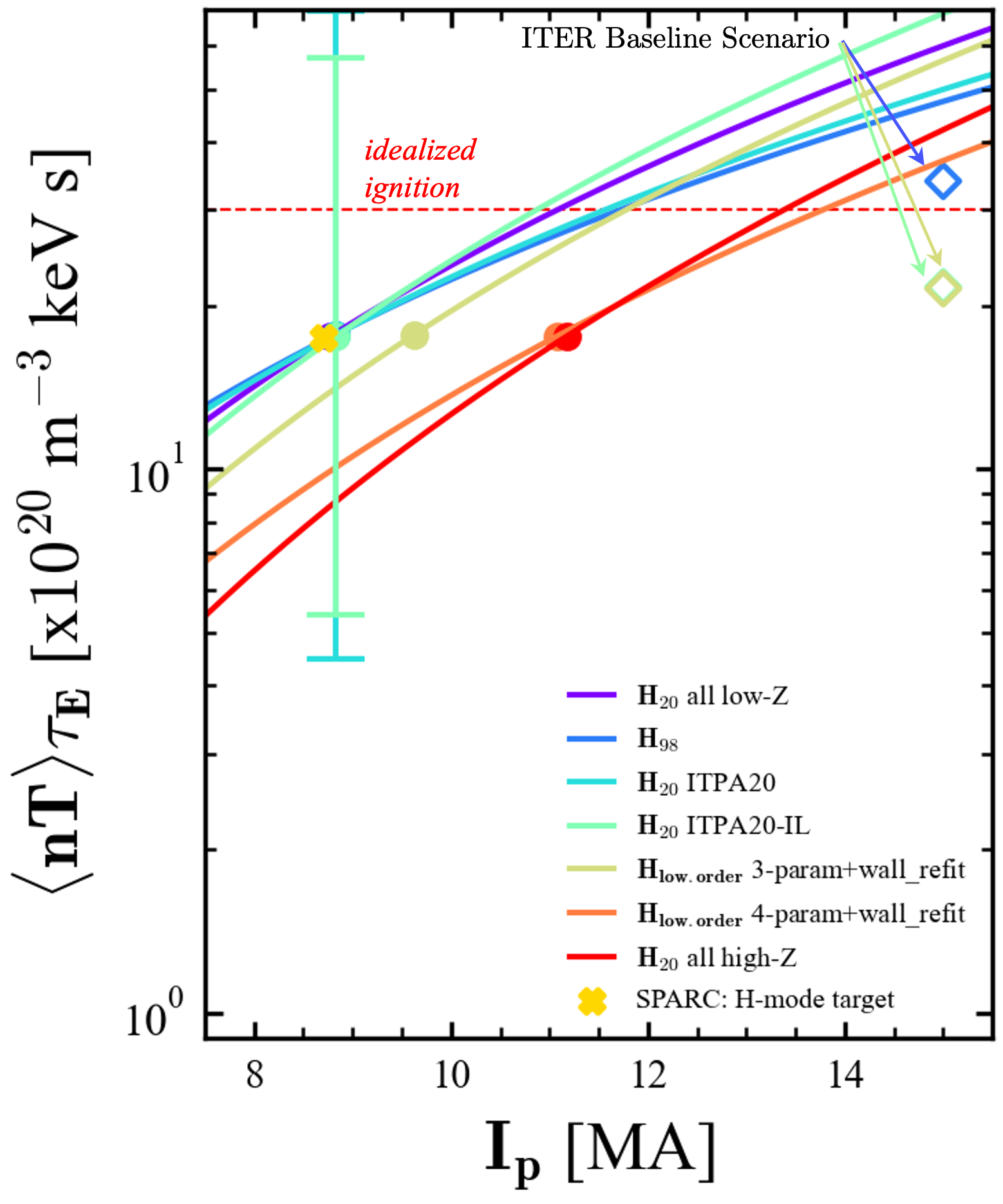}
    \end{subfigure}
    \captionsetup{width=\linewidth, font= footnotesize}
    \caption{ $\langle nT \rangle \tau_E$ for the SPARC H-mode scenario for an ensemble of confinement time scalings, including the design-point, $\tau_{E_{98,y2}}$. The 15MA ITER Baseline Scenario projected with the optimal low order scaling (developed herein), IPB98(y,2), and ITPA20-IL are shown for reference.}
    \label{fig: <nT>tau ensemble}
\end{figure}

\noindent Applying the optimal low order scaling to the SPARC H-mode reference\cite{creely_overview_2020} as a test case yields confinement times below those predicted by IPB98(y,2). An increase of at least 10\% on the required $I_p$ over the IPB98(y,2) prediction is noted for the optimal low-order scalings. Some qualitative observations using a simple pre-burn power balance model are documented in figure  \ref{fig: <nT>tau ensemble} and tables \ref{table: tauE shortlisted survey} and \ref{table: <nTtau> correspondence:  Ip for Q=10}. 

\begin{table}[h]
  \centering
  \begin{tabular}{|l||c|c|c|}
    \hline
    \textbf{Scaling} & $\boldsymbol{\tau_E}$ \textbf{[s]} & $H_{98,y2}$ & $\langle n T\rangle \tau_E  \Bigr |_{\propto H_{98,y2}^2}$   \\%& $\boldsymbol{Q}$ \\
    \hline
    IPB98(y,2)                          & 0.77 & 1.0  & 18\\%& 9.7 \\
    ITPA20                              & 0.77 & 1.0  & 18\\%& 9.5 \\
    ITPA20: all low-$Z$                 & 0.77 & 1.0  & 18\\%& 9.6 \\
    ITPA20: all high-$Z$                & 0.52 & 0.69 & 9 \\%& 2.4 \\
    Optimal low order fit (N=3)         & 0.68 & 0.88 & 14 \\%& 2.2 \\
    $\sim$Optimal low order fit ($N=4$) & 0.58 & 0.75 & 10  \\%& 2.2 \\
    \hline
  \end{tabular}
\captionsetup{width=\linewidth, font= footnotesize}
\caption{Survey of confinement time scalings and their predictions for the SPARC H-mode scenario. This expression of $\langle n T \rangle \tau |_{\propto H_{98}^2}$ is in units of $10^{20}$ keV $m^{-3}$s and is calculated using its proportionality with $H^2$; this is distinct from the explicit evaluation of $\langle nT\rangle \tau$ presented in figure \ref{fig: <nT>tau ensemble}. The reference $\langle nT \rangle \tau_E$ assumes parabolic profiles.}
\label{table: tauE shortlisted survey}   
\end{table}

 The values of $\langle nT\rangle \tau$ presented in figure \ref{fig: <nT>tau ensemble} and table \ref{table: tauE shortlisted survey} consider variation of $I_p$ only, since it is found to be the first-order important engineering variable at fixed geometry. However, for a given machine, the achievable $I_p$ is at least limited by low-$q$ MHD instabilities. Achieving the scenario reference $\langle nT \rangle \tau$ for the minimally adequate scaling requires approximately $I_p=9.6$MA, while its incrementally more complex neighbor $N=4$ requires $I_p=11.1$MA. While we do not find $\langle nT\rangle \tau$ to have any direct dependence on the toroidal field strength, operating at this plasma current without a change in geometry requires $B_T \approx 11.9-13.8$T to operate at $q_{95}\approx3.0$, the lower end of which is achievable in the SPARC design. Operation at $I_p=11.1$MA at maximum field $B_T=12.2$T yields $q_{95} \approx2.66$, which is approaching dangerous low-$q$ MHD territory. We can consider how $\langle nT \rangle \tau$ varies with $I_p$ and $B_T$ with $q_{95}$ fixed at the design point by varying $B_T$ proportionally with $I_p$, presented in table \ref{table: <nTtau> correspondence:  Ip for Q=10} and figure \ref{fig: <nTtau> ensemble with fixed q=3.4}. The net effect of fixed operation at $q_{95}=3.4$ on $\langle nT \rangle \tau$ is that it is unchanged for the low-order scalings, as $B_T$ is not present; modestly higher for scalings where $\alpha_{B_T}>0$ (e.g. IPB98(y,2)); and moderately lower where $\alpha_{B_T}<0$ (e.g. ITPA20-IL).  Furthermore, we note the relatively advantageous position that SPARC sits in through its lower loss power. Increasing $I_p$ up to ITER order could yield an $\langle nT \rangle \tau_E$ up to two to three times greater than that of the ITER Baseline Scenario.

\begin{table}
\centering
\begin{tabular}{|l||c|c|}
\hline
\textbf{Scaling} (as named in fig \ref{fig: <nT>tau ensemble}) & \textbf{$I_p$ [MA] for $Q\approx 11$} & $B_T \Big|^{q_{95}=3.4}_{I_p  \text{ for } Q_{ref}}$  [T]\\
\hline
$H_{20}$ all low-Z & 8.7 & 12.2\\
$H_{98}$ & 8.7 & 12.2\\
$H_{20}$ ITPA20 & 8.7 & 12.2 \\
$H_{20}$ ITPA20-IL & 8.7 & 12.2\\
$H_{\text{low. order}}$ 3-param+wall\_refit & 9.6 & 13.5  \\
$H_{\text{low. order}}$ 4-param+wall\_refit & 11.1 & 15.5\\
$H_{20}$ all high-Z & 11.2 & 17.0\\
\hline
\end{tabular}
\captionsetup{width=\linewidth, font= footnotesize}
\caption{$I_p$ required to achieve the reference $Q$ set out in the SPARC H-mode scenario for an ensemble of confinement scalings. Accompaniment to figure \ref{fig: <nT>tau ensemble}. The values for $B_T \Big|^{q_{95}=3.4}_{I_p  \text{ for } Q_{ref}}$ are taken from appendix figure \ref{fig: <nTtau> ensemble with fixed q=3.4}.}
\label{table: <nTtau> correspondence:  Ip for Q=10}
\end{table}

\begin{figure}[ht!]
    \centering
    \includegraphics[width= 0.99\linewidth]{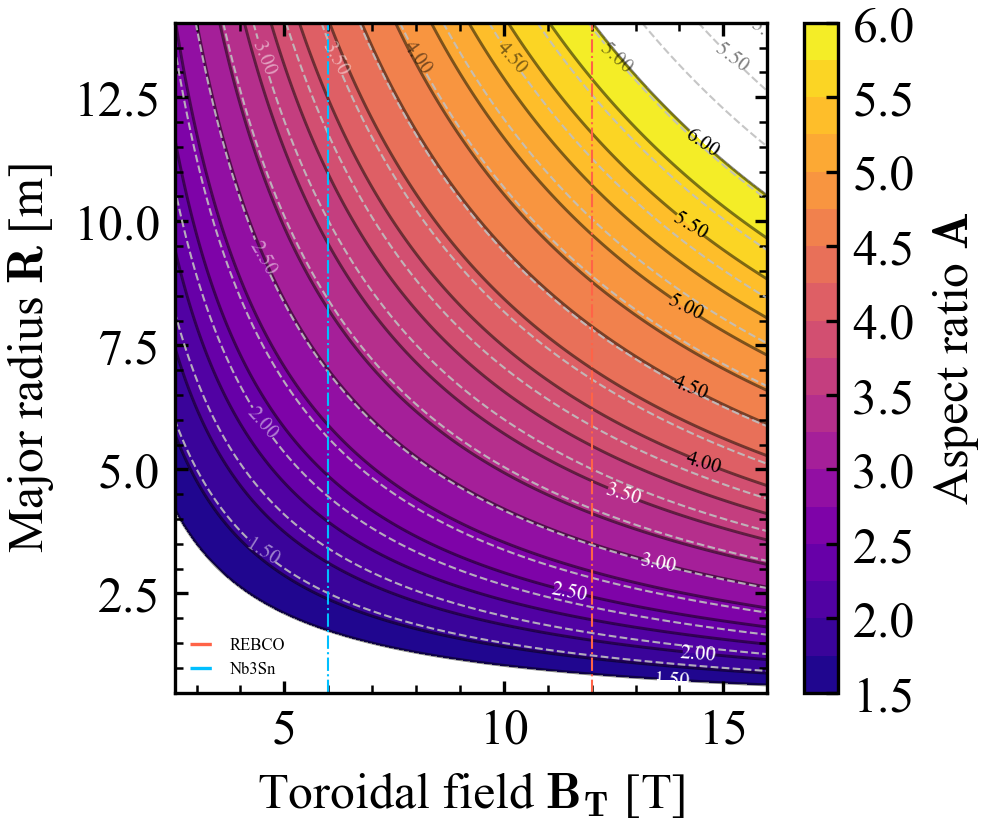}
    \captionsetup{width=\linewidth, font= footnotesize}
    \caption{Machine size for safety factor $q^*_{\text{Uckan}}=3.1$ as a function of $B_T$ and aspect ratio for a 1 GWth class reactor. Black contour lines correspond to the all-machine-scaling ($\propto I_p^{2.30}$), with contour labels (black or white) on the higher field side of the plot, and the dashed-gray contours correspond to the achieved/D-T-equivalent-scaling ($\propto I_p^{1.98}$), contour labels on the lower field side. The vertical \texttt{-.} lines correspond to the order of on-axis field achievable by Nb3Sn and REBCO tape magnet technologies.}
    \label{fig: 1GW class reactor contours}
\end{figure}

Now we explore some back-of-the-envelope implications of the empirical fusion power scaling derived in the previous section. Prototype power plants generally aim for $Q_p \geq 10$ and $Q_e > 1$. Achieving such fusion gain is an essential milestone, and remains a major scientific challenge for the next generation of reactor-scale devices. However, high gain is a necessary condition for a reactor, not a sufficient one. Even in the idealized limit of very large scientific gain, commercial relevance requires sufficient fusion power production. In practice, a reactor must deliver several hundred megawatts of net electricity, and most contemporary reactor studies indicate that this in turn requires gigawatt-class fusion power\cite{luxCommercializationFusionPower2022}. The single-parameter fusion power scaling indicates that achieving 1 GW of fusion power requires a plasma current of order $I_p \sim 20$--$30$MA. We then take the maximum on-axis toroidal field that can be produced by either HTS or LTS technology, and finally determine the machine size required to stably accommodate such a plasma. Assuming HTS technology can support a peak TF-coil field of $\sim 20$T, corresponding to roughly $B_T \sim 12$T on axis, and requiring operation at $q^*_{\text{Uckan}}=3.1$ to remain above MHD limits, we use  $q^*_{\text{Uckan}} = 5 \frac{\left (1+\kappa^2(1+2\delta^2 - 1.2\delta^3) \right)}{2} \frac{a^2}{R} \frac{B_T}{I_p}$ to relate critical design quantities. For a specified plasma current, toroidal magnetic field, aspect ratio, and elongation, this relation determines the machine size needed to carry the required current at sufficiently high safety factor. Taking representative ITER- and SPARC- like shaping, we find that a machine with major radius of order $R\sim 4$m is sufficient to support the required plasma current in the REBCO enabled field space. A contour plot of machine size, toroidal field, and aspect ratio for a 1 GW class tokamak using the empirical fusion power scalings proposed in this work is presented in figure \ref{fig: 1GW class reactor contours}. A 0.5 GW machine at ITER's shaping ($A=3.1$, $\kappa=1.85$, $\delta=0.48$) and HTS magnets would require $I_p\approx 15$ MA and geometric size of approximately $R\approx 2.8$ m; using LTS magnets at the ITER level would require $R\approx6.3$ m, which is within 2\% of ITER's final design value of $6.2$ m. Scaling up to 2 GW with HTS magnets requires $R\approx 5$ m, whereas LTS requires $R\approx 10$ m, which is within 10\% of the EU-DEMO design of $R=9$ m.

HTS enables reactor-scale $I_p$ to be carried in a smaller device, rather than by eliminating the need for high plasma current in the first place. This back-of-the-envelope estimate should, however, be interpreted only as a crude safety-factor-based lower bound on machine size, since it neglects additional requirements associated with tritium breeding, shielding, stresses, heat exhaust, and other reactor engineering constraints.

\section{Discussion and outlook}
Our empirical analysis suggests that, for conventional aspect-ratio tokamaks in the auxiliary-heated operating space examined here, plasma current appears as the most direct engineering lever on fusion performance, while toroidal field enters chiefly through the equilibrium and stability constraints required to access and sustain that high-current state. In that sense, the decades-long stagnation in tokamak fusion performance can be viewed as a stagnation in experimentally explorable $I_p$. Since the record fusion power set in 1997 by JET, the accessible performance space has remained effectively frozen, with much of the community's innovation shifted toward advanced scenarios, stable high-$\beta$ operation, low-aspect-ratio configurations, and magnet technology development.

This high-level conclusion is not new. Wesson framed reactor progress in these terms 27 years ago, emphasizing the dominant role of plasma current and noting that a JET-like machine would need a plasma current of order $20$ MA to reach Lawson-level ignition conditions\cite{wesson_science_1999}. That figure is, notably, close to the range the multi-machine empirical projection presented in section \ref{Section:  Empirical scaling for fusion power ($P_{fus}$):  establishing an empirical yard-stick for progress in fusion power} suggests for a gigawatt-class reactor.

%%-- post-word smithed
This fusion power scaling differs in form from the more frequently written scaling in terms of $B_T^4$. However, the scaling found here is not inconsistent with that form for kink-stable operation. With average plasma current density fixed, and where $q \propto \frac{\kappa a^2 B_T}{R I_p}$, the $I_p$-based fusion-power scaling can be recast in the form $P_{\mathrm{fus}} \propto B_T^4$. The distinction suggested by the present empirical picture concerns the role played by $B_T$: plasma current emerges as the most direct engineering lever on fusion performance, while higher toroidal field is required to stabilize and sustain that high-current state. This framing also helps place statements that confinement strongly improves with magnetic field at fixed safety factor\cite{sorbomARCCompactHighfield2015} in a more precise empirical context.

In this vein, one may also reinterpret a commonly quoted criterion for burning plasmas, $B_T R \gtrsim 20\ \mathrm{T\cdot m}$  \cite{hutchinson_feasibility_2022}. Higher field and larger size can trade against each other in accessing and sustaining high performance at acceptable stability margins \cite{zohm_size_2019}. The quantity $B_T R$ can be interpreted to arise through the equilibrium constraints that connect it to plasma current. For fixed safety factor, aspect ratio, and shaping, the $B_T R$ condition can be written as a minimum-current requirement. Using ITER-like shaping, the condition $B_T R = 20\ \mathrm{T\cdot m}$ corresponds to approximately $I_p = 9\ \mathrm{MA}$. In that sense, and with the perspective of the empirical analysis presented in this work, the typical $B_T R$ criterion may be viewed as a proxy for the plasma current needed to reach reactor-scale performance at acceptable MHD stability margin.

The advent of scalable HTS technology materially changes the design landscape. HTS enables high-field coils that can support the high-$I_p$ operation demanded by empirical scalings for strong fusion performance at substantially smaller size than would be required with conventional LTS. The empirical picture indicates that HTS does not eliminate the need for high plasma current; rather, it makes reactor-relevant plasma current more accessible in compact devices. This is the sense in which the compact high-field path appears especially attractive.

More specifically, the confinement-time analysis presented here indicates that low-order empirical scalings, centered near $N=3$ to $N=4$, provide an appropriate basis for extrapolation. These models recover familiar qualitative insight: plasma current, machine size, and heating power are the dominant engineering levers on confinement, while several other dependencies remain substantially uncertain in the updated database. In addition, empirically treating the transition to metal walls as a simple multiplicative confinement penalty yields roughly a 10--15\% degradation. When applied to reactor-relevant operating space, these low-order scalings imply a roughly 12--25\% reduction in expected confinement relative to classic projections. The projections from these extrapolation-forward scalings are not definitive in any sense, nor impossible to outperform. However, we believe the $N=3$, $N=4$ projections provide a more appropriate nominal baseline against which reactor designs can be evaluated.

A related implication concerns non-H-mode operation. While international confinement data centralization increasingly focused on the favorable H-mode regime, a non-H-mode path may still remain attractive for some stages of FPP operation because it offers an operationally simpler route to fusion gain milestones. Recent integrated modeling of projected SPARC plasmas \cite{rodriguez-fernandez_core_2024} shows that $Q>1$ is obtained only for the upper range of a C-Mod-informed set of the highest-performing non-H-mode edge conditions, requiring edge pressures that are a substantial fraction of H-mode values. The underlying non-H-mode database does not exclude I-mode, and explicitly notes that the highest temperature edges are likely associated with I-mode-like conditions. Breakeven may therefore be accessible only if the machine reaches such favorable non-H-mode edge performance. Achieving breakeven in sub- H-mode conditions remains a daunting task for any machine, especially in light of this recent work. That risk can be mitigated by designing machines to access the highest achievable confinement.

The design implication that follows is straightforward. If $I_p$ is the principal performance actuator, then reactors should be designed to accommodate higher-$I_p$ operation in order to achieve fusion performance targets historically projected using empirical confinement scalings, especially for metal-walled machines. In that light, high-$I_p$ reactor concepts, including STEP-SPP2 \cite{meyerPlasmaBurnMind2024} and EU-DEMO \cite{federiciDEMODesignActivity2018}, appear especially attractive, before considering the many other critical optimizations necessary for reactor design.

\clearpage
\section{Appendix}
\subsection{Full $N$-parameter models' details}
\label{appendix: Full $N$-parameter models' details}
Full details of the $N$-parameter models are presented in tables \ref{table: incremental_scalings_raw} and  \ref{table: incremental_scalings_aug}. Table \ref{table: incremental_scalings_raw} presents the WLS fit without explicit wall-discriminating consideration, while \ref{table: incremental_scalings_aug} presents the regression with the wall-dependent multiplier, where metal wall confinement times are multiplied by WallConst.

% ------------------------- RAW (no WallConstant) -------------------------
\begin{table*}[htbp]
\centering
\begin{tabular}{|l||c|c|c|c|c|c|c|c|c|c||c|}
\hline
\textbf{Model}
& $C_1$
& $I_p$
& $R_{\mathrm{geo}}$
& $P_{\mathrm{l,th}}$
& $\kappa$
& $\bar{n}_{e}$
& $M_{\mathrm{eff}}$
& $(1+\delta)$
& $B_T$
& $\epsilon$
& $R^2$ \\
\hline

1-param
& 0.1005
& 1.1758
& ---
& ---
& ---
& ---
& ---
& ---
& ---
& ---
& 0.7770 \\

2-param
& 0.0584
& 0.8326
& 0.8791
& ---
& ---
& ---
& ---
& ---
& ---
& ---
& 0.8560 \\

3-param
& 0.1244
& 1.3194
& 1.0665
& -0.5230
& ---
& ---
& ---
& ---
& ---
& ---
& 0.9352 \\

4-param
& 0.0931
& 1.2293
& 1.2280
& -0.5676
& 0.6215
& ---
& ---
& ---
& ---
& ---
& 0.9433 \\

5-param
& 0.0551
& 1.1758
& 1.4935
& -0.6445
& 0.6917
& 0.2711
& ---
& ---
& ---
& ---
& 0.9526 \\

6-param
& 0.0486
& 1.1479
& 1.5151
& -0.6433
& 0.6710
& 0.2573
& 0.2269
& ---
& ---
& ---
& 0.9549 \\

7-param
& 0.0460
& 1.1340
& 1.5532
& -0.6478
& 0.5641
& 0.2666
& 0.2454
& 0.2547
& ---
& ---
& 0.9555 \\

8-param
& 0.0467
& 1.1177
& 1.4908
& -0.6432
& 0.6571
& 0.2044
& 0.2220
& 0.3760
& 0.1062
& ---
& 0.9567 \\

9-param
& 0.0531
& 0.9777
& 1.7045
& -0.6679
& 0.8077
& 0.2439
& 0.1985
& 0.3457
& 0.2163
& 0.3475
& 0.9566 \\

\hline
\end{tabular}
\captionsetup{width=\linewidth, font= footnotesize}
\caption{Highest $R^2$ model for each $N$-parameter model. No explicit treatment of the wall material is made in these fits.}
\label{table: incremental_scalings_raw}
\end{table*}

% ------------------------- AUG (with WallConstant) -------------------------
\begin{table*}[htbp]
\centering
\begin{tabular}{|l||c|c|c|c|c|c|c|c|c|c|c||c|c|}
\hline
\textbf{Model}
& $C_1$
& $I_p$
& $R_{\mathrm{geo}}$
& $P_{\mathrm{l,th}}$
& $\kappa$
& $\bar{n}_{e}$
& $M_{\mathrm{eff}}$
& $(1+\delta)$
& $B_T$
& $\epsilon$
& WallConst.
& $R^2$ \\
\hline

1-param+wall
& 0.1079
& 1.2175
& ---
& ---
& ---
& ---
& ---
& ---
& ---
& ---
& 0.7620
& 0.8022 \\

2-param+wall
& 0.0627
& 0.8750
& 0.8718
& ---
& ---
& ---
& ---
& ---
& ---
& ---
& 0.7725
& 0.8800 \\

\rowcolor{gray!15}
3-param+wall
& 0.1226
& 1.3082
& 1.0521
& -0.4915
& ---
& ---
& ---
& ---
& ---
& ---
& 0.8958
& 0.9393 \\

4-param+wall
& 0.0934
& 1.2252
& 1.2086
& -0.5410
& 0.5900
& ---
& ---
& ---
& ---
& ---
& 0.9184
& 0.9464 \\

5-param+wall
& 0.0508
& 1.1593
& 1.5022
& -0.6080
& 0.6450
& 0.3166
& ---
& ---
& ---
& ---
& 0.8537
& 0.9583 \\

6-param+wall
& 0.0463
& 1.1383
& 1.5186
& -0.6090
& 0.6314
& 0.3034
& 0.1777
& ---
& ---
& ---
& 0.8611
& 0.9598 \\

7-param+wall
& 0.0437
& 1.1243
& 1.5569
& -0.6135
& 0.5239
& 0.3126
& 0.1963
& 0.2557
& ---
& ---
& 0.8611
& 0.9603 \\

8-param+wall
& 0.0444
& 1.1060
& 1.4880
& -0.6074
& 0.6256
& 0.2453
& 0.1689
& 0.3899
& 0.1175
& ---
& 0.8571
& 0.9617 \\

9-param+wall
& 0.0531
& 0.9051
& 1.7925
& -0.6388
& 0.8370
& 0.3060
& 0.1297
& 0.3482
& 0.2756
& 0.4956
& 0.8433
& 0.9624 \\

\hline
\end{tabular}
\caption{Highest $R^2$ model for each $N$-parameter model with wall-discriminating constant natively fitted.}
\label{table: incremental_scalings_aug}
\end{table*}

\begin{table*}[htbp]
\centering
\renewcommand{\arraystretch}{1.40}
\begin{tabular}{|l||c|c|c|c|c|c|c|c|c|c|}
\hline
\textbf{Model}
& $C_1$
& $I_p$
& $R_{\mathrm{geo}}$
& $P_{\mathrm{l,th}}$
& $\kappa$
& $\bar{n}_{e}$
& $M_{\mathrm{eff}}$
& $(1+\delta)$
& $B_T$
& $\epsilon$ \\
\hline

1-param
& ${}^{+0.013}_{-0.012}$
& $\pm 0.17$
& ---
& ---
& ---
& ---
& ---
& ---
& ---
& --- \\

2-param
& ${}^{+0.020}_{-0.014}$
& $\pm 0.18$
& $\pm 0.42$
& ---
& ---
& ---
& ---
& ---
& ---
& --- \\

\rowcolor{gray!15}
3-param
& ${}^{+0.048}_{-0.038}$
& $\pm 0.15$
& $\pm 0.24$
& $\pm 0.12$
& ---
& ---
& ---
& ---
& ---
& --- \\

4-param
& ${}^{+0.048}_{-0.035}$
& $\pm 0.17$
& $\pm 0.25$
& $\pm 0.13$
& $\pm 0.44$
& ---
& ---
& ---
& ---
& --- \\

5-param
& ${}^{+0.046}_{-0.027}$
& $\pm 0.17$
& $\pm 0.38$
& $\pm 0.13$
& $\pm 0.45$
& $\pm 0.17$
& ---
& ---
& ---
& --- \\

\hline
\end{tabular}
\renewcommand{\arraystretch}{1.0}
\caption{RBAYES-derived practical uncertainty estimates for the scaling exponents of the $N=1$ to $N=5$ scalings without wall discrimination (table \ref{table: incremental_scalings_raw}).}
\label{table: incremental_scalings_aug_uncertainties}
\end{table*}

\subsection{Model performance with increasing model complexity $N\rightarrow9$}

\begin{figure}[ht!]
    \centering
    \includegraphics[width=0.99\linewidth]{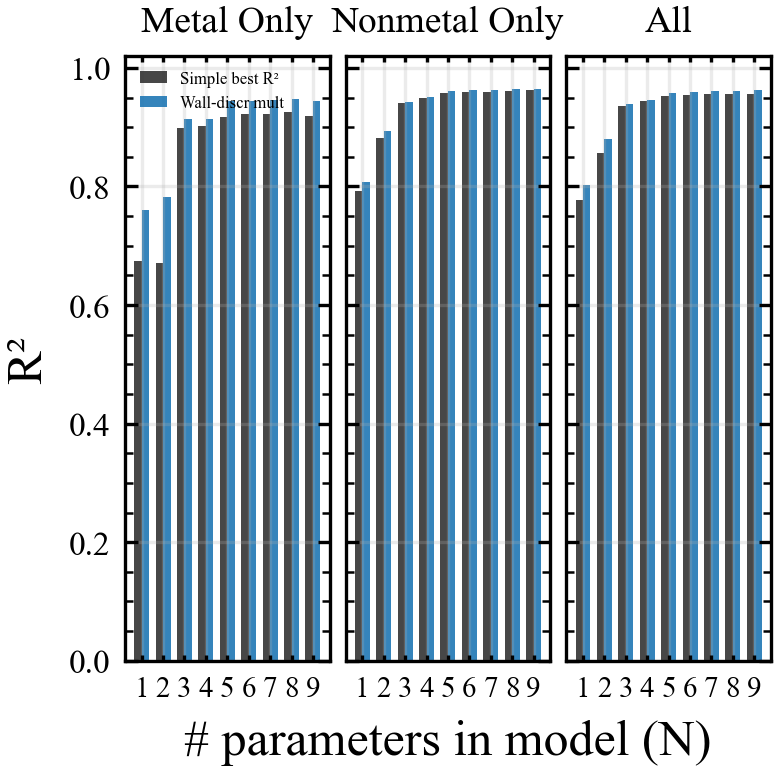}
    \captionsetup{width=\linewidth, font= footnotesize}
    \caption{Model $R^2$ partitioned by wall as a function of model complexity. The non-augmented N=9 model has a DB5.2.3-STD5 $R^2=0.9567$, while the N=3 model has $R^2=0.9353$. The addition of six more parameters yielded an improvement of $\Delta R^2=0.0214$, which amounts to an $R^2$ improvement per additional parameter of $\Delta R^2/\Delta N_{3-9} = 0.35\%$.}%R^2=0.9624$, which has an $0.9394$  }
    \label{fig: R^2 by wall-partition as up to N=9}
\end{figure}

\noindent As model complexity increases, $R^2$ for the full database increases. The $R^2$ for each wall subset (low-Z:  "metal only", high-Z:  "nonmetal only") is presented in figure \ref{fig: R^2 by wall-partition as up to N=9}.

\begin{figure}[ht!]
    \centering
    \includegraphics[width=0.99\linewidth]{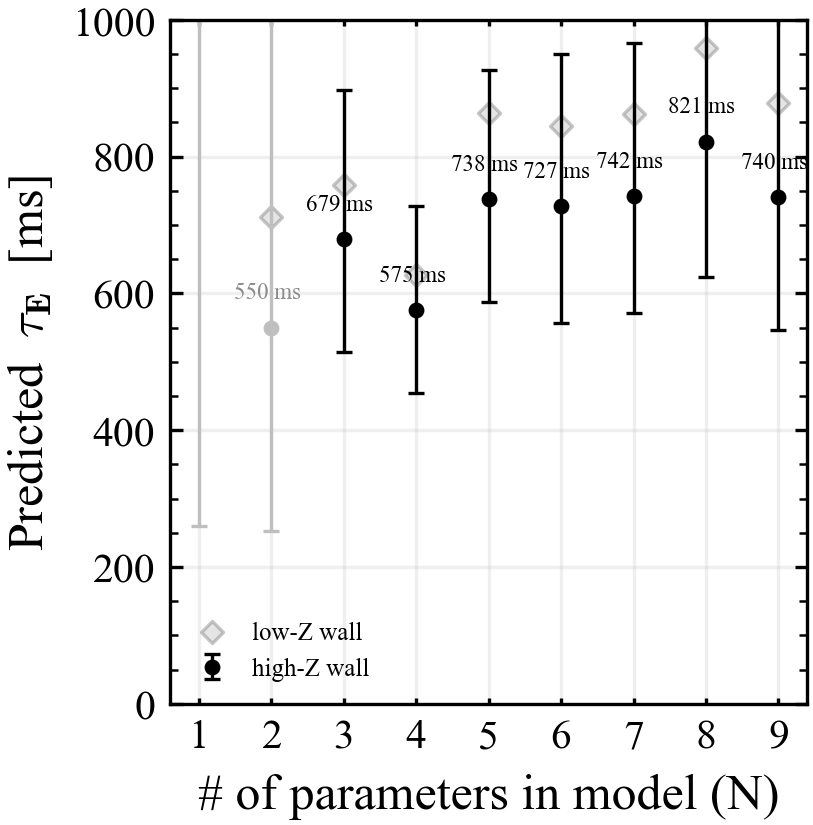}
    \captionsetup{width=\linewidth, font= footnotesize}
    \caption{$\tau_E$ predictions for SPARC's reference H-mode scenario for models $N=1-9$ for three different model types. The filled circles represent predictions from the wall-discriminating scalings using a high-Z wall from table \ref{table: incremental_scalings_aug}; the open diamonds represent predictions from the same scalings but using a low-Z wall. The error bars shown are $\pm E_{far}$, which we only present for illustrative purposes.}
    \label{fig: tauE+EBs for N=1-9}
\end{figure}

The high-Z confinement time prediction for the SPARC H-mode reference scenario is presented in figure \ref{fig: tauE+EBs for N=1-9}. The error bars presented in figure \ref{fig: tauE+EBs for N=1-9} indicate the average $N$-parameter model scatter projected from the extrapolation error metric $E_{far}$ exhibited by a range of priors, introduced in appendix section \ref{appendix: Systematic determination of $N-$parameter model noninferiority to highest complexity model}. The far-tail error is projected linearly in log-space, with a distance metric of $d = \ln \left(\frac{\tau_E}{\tau_{E,b}}\right)$, where $\tau_{E,b}$ is the priors-specific boundary. We only include the $E_{far}$-based error bars for illustrative purposes. A quotable systematic uncertainty estimation is the robustified bayesian (RBAYES) error bars presented in table \ref{table: incremental_scalings_aug_uncertainties}, introduced by Verdoolaege et al \cite{verdoolaege_updated_2021}.

\subsection{Systematic analysis for optimal model complexity truncation}
\subsubsection{Systematic determination of $N-$parameter model noninferiority to highest complexity model on a low-$\tau_E$ to high-$\tau_E$ basis}
\label{appendix: Systematic determination of $N-$parameter model noninferiority to highest complexity model}

\noindent Before carrying out the simulated extrapolation error inflation exercise, we first split the standard set of the ITPA H-mode confinement database in $\tau_E$ by distribution percentile. We built N-parameter models for incrementing subsets of the $\tau_E$ distribution ranging $\tau_{E} \in [0, 30-60]$ percentile. Specifically, we selected ten upper training boundaries $q_{\mathrm{train}}$ uniformly spanning the interval $0.30 \le q_{\mathrm{train}} \le 0.60$, and for each such value constructed a training set consisting of all database entries with $\tau_E$ percentile less than or equal to $q_{\mathrm{train}}$. We refer to the resulting ensemble of $N$-parameter models constructed from each training subset split as an empirically motivated prior ranking for the extrapolation task. We evaluate each $N$-parameter candidate model’s performance on the far-tail extrapolation range spanning the 60th to 95th percentile. In this way, each fit was intentionally restricted to the lower-confinement portion of the database and then evaluated on progressively higher-confinement data that lay outside the support of the training distribution. For each training subset and each candidate model order $N$, we refit the $N$-parameter model form selected from the full-database analysis using the same weighted least-squares methodology. In this way, the extrapolation study evaluates, at each complexity order, the full-database-selected model form after calibration on the restricted training subset.

The extrapolation region was resolved into percentile shells of fixed width $\Delta q = 0.05$, beginning just above the training boundary and extending to the 95th percentile of the $\tau_E$ distribution. We computed log-space prediction errors for each extrapolation shell, and in particular a far-tail metric $E_{\mathrm{far}}$, defined as the log-space RMSE over the last 10\% of the test range, corresponding here to the $85$th--$95$th percentiles of the $\tau_E$ distribution. This quantity was adopted as the principal figure of merit for judging whether a model preserved acceptable predictive fidelity deep in the extrapolative regime.

The resulting extrapolation error curves over this simulated extrapolation range averaged over each priors set are presented in figure \ref{fig: extrapolation in the far-tail, averaged over priors} for each $N$-parameter model. The extrapolation error curves show a clear separation between the lowest-order and higher-order models. The $N=1$ and $N=2$ models exhibit marked error inflation as evaluation moves away from the upper boundary of their training support and into the far tail of the held-out distribution, whereas the $N \ge 3$ models show substantially more stable behavior. On this basis of inadequacy, the one- and two-parameter models were excluded from detailed consideration in the remainder of the manuscript.

\begin{figure}
    \centering
    \includegraphics[width=0.99\linewidth]{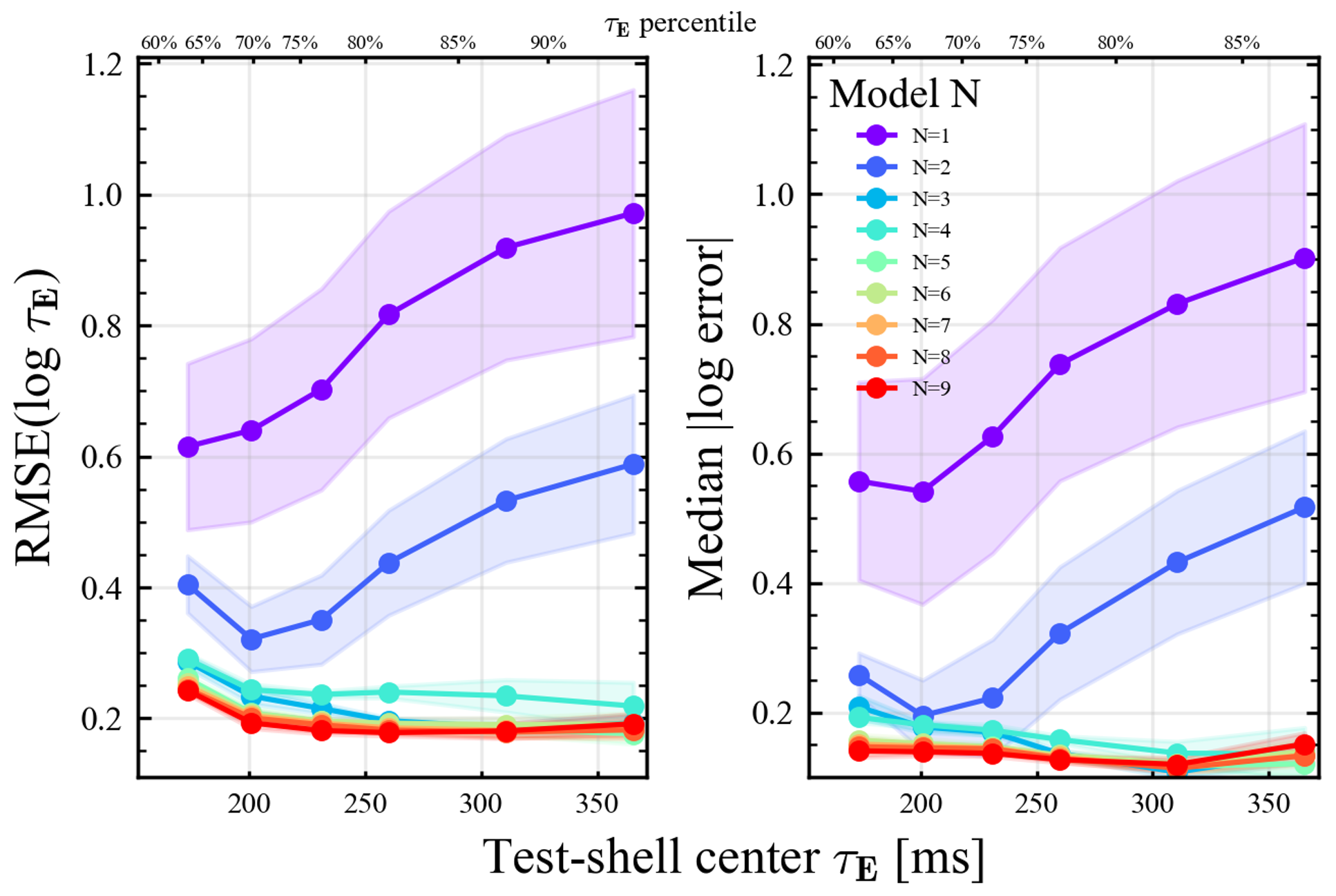}
    \captionsetup{width=\linewidth, font= footnotesize}
    \caption{RMSE and median absolute error of $\ln \tau_E$, averaged over the lower range of $\tau_E$ spanned by the priors, and plotted as a function of the simulated "out-of-distribution" $\tau_E$. \textit{Test-shell} refers to the discrete bands in the simulated extrapolation range of $\tau_E$ over which the errors were evaluated, for which the center in distribution space serves as the abscissa.}
    \label{fig: extrapolation in the far-tail, averaged over priors}
\end{figure}

\begin{figure}[ht!]
    \centering
    \includegraphics[width=0.99\linewidth]{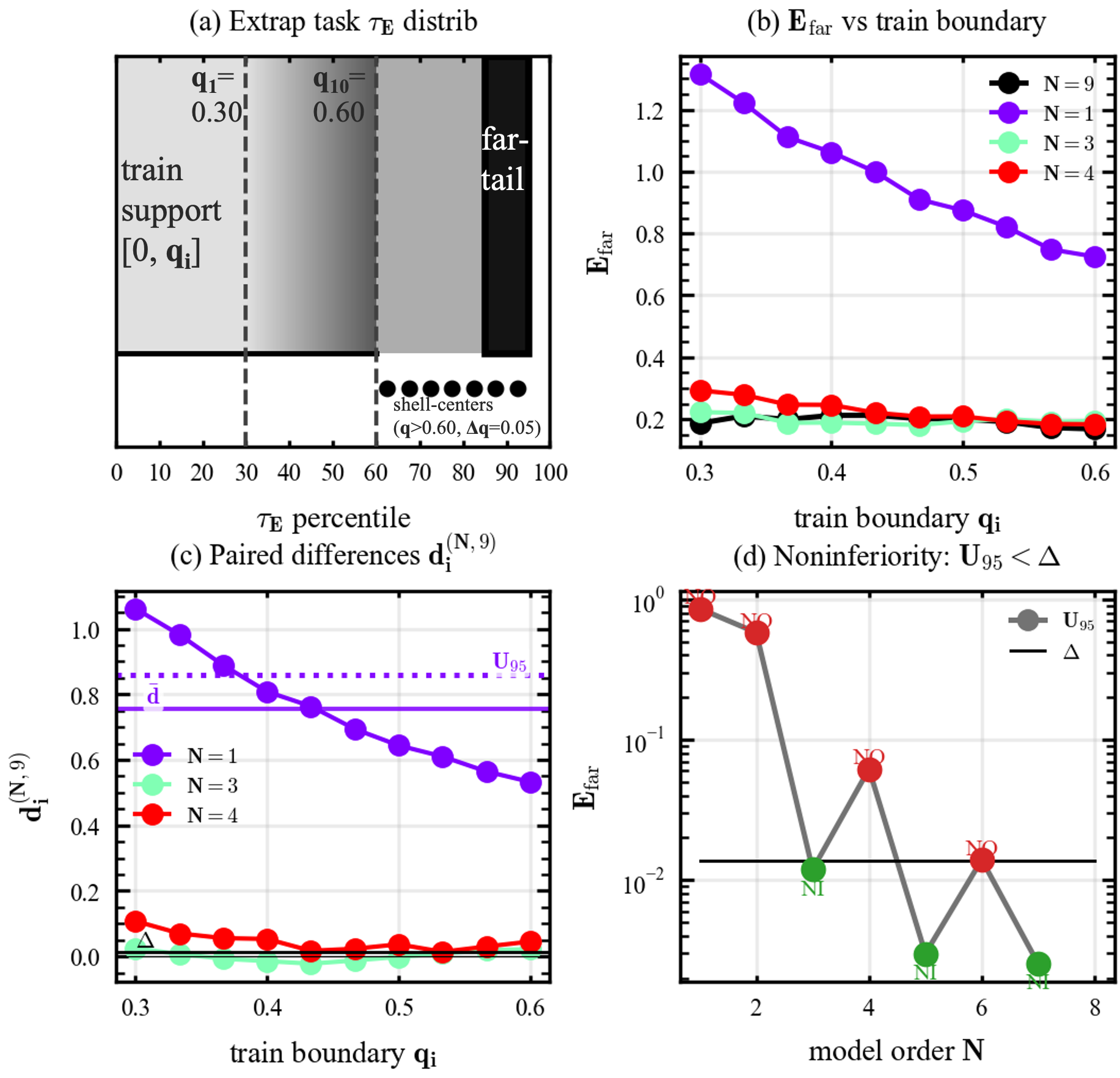}
    \captionsetup{width=\linewidth, font= footnotesize}
\caption{
    Schematic of the procedure used for the percentile-shell noninferiority analysis. 
    (a) For each training boundary $q_i$, a model of order $N$ is fit only on the lower-$\tau_E$ support $[0,q_i]$ and then evaluated on the held-out extrapolation region above $q_i$, with the far-tail metric $E_{\mathrm{far}}$ defined over the $85$th--$95$th percentile range. For each model order $N$, a distinct set of model exponents is fit on each training-support range $[0,q_i]$.
    (b) The resulting $E_{\mathrm{far}}^{(N)}(q_i)$ curves are compared against the reference $N=9$ model. 
    (c) For each $q_i$, the paired difference $d_i^{(N,9)} = E^{(N)}(q_i)-E^{(9)}(q_i)$ is formed, from which $\bar d$ and the one-sided 95\% upper confidence bound $U_{95}$ are computed. 
    (d) Noninferiority is declared when $U_{95}<\Delta$, where $\Delta = \mathrm{SD}(E^{(9)}(q_i))$ is a measure of the reference-model variation across training boundaries.
    }
    \label{fig: noninferiority analysis procedure}
\end{figure}

To determine whether the lower-order models were meaningfully worse than the highest-complexity model, we performed a formal noninferiority analysis using the $N=9$ model as the reference. For each training boundary $q_i$ and model order $N$, we computed the paired difference in the far-tail error metric,
$$
d_i^{(N,9)} = E^{(N)}(q_i) - E^{(9)}(q_i),
$$
so that positive values correspond to worse extrapolation performance of the lower-order model. The comparison is paired in the sense that both errors are evaluated at the same training boundary $q_i$. For each model order $N$, we estimated the mean paired difference $\bar d^{(N,9)}$, its standard error, and the one-sided 95\% upper confidence bound, 
$$
U_{95}(\bar{d}^{(N,9)}) = \bar{d}^{(N,9)} + t_{0.95,\nu}\,\mathrm{SE}(\bar d^{(N,9)}),
$$
with $\nu = n_{\mathrm{pairs}}-1$. The noninferiority margin was defined as the standard deviation of the corresponding $N=9$ far-tail extrapolation error $E^{(9)}$ across the set of training boundaries, $\Delta = \mathrm{SD}\!\left(E^{(9)}(q_i)\right)$. A lower-order model was declared noninferior whenever $U_{95}(\bar{d}^{(N,9)}) < \Delta$. A schematic of the intermediate steps and key quantities in the noninferiority analysis is presented in figure \ref{fig: noninferiority analysis procedure}. This criterion asks whether any possible excess extrapolation error of a lower-order model is smaller than the intrinsic variation in the reference model's own extrapolation performance as the training-support cutoff is varied. 

Under this criterion, the $N=3$ model was noninferior to the $N=9$ model for the far-tail metric $E_{\mathrm{far}}$, i.e. for prediction in the $85$th--$95$th percentile region of $\tau_E$. Combined with the knee observed in the gain of held-out variance capture with increasing model order, this result supports the interpretation that three parameters constitute the lowest-complexity model class that reproduces the extrapolative behavior of the highest-complexity model without meaningful degradation by the present criterion, at least within the simulated extrapolation task permitted by the database. We therefore adopt the $N=3$ model as the principal low-order confinement scaling for the purposes of this study. Although the $N=4$ and $N=6$ models remain within the broader stable-extrapolation cluster, they did not satisfy the present noninferiority criterion relative to the $N=9$ model in $E_{\mathrm{far}}$. Taken together, we do not interpret these results to conclude that the $N=4$ and $N=6$ models perform poorly in an absolute sense, only that they do not satisfy the specific noninferiority criterion for this specific extrapolation task. This is apparent in figure \ref{fig: noninferiority analysis procedure}'s panels c) and d).

\subsubsection{Machine-organized extrapolation performance: chronology, $\tau_E$, and composite reactorwardness}
The preceding appendix subsection asked whether a model fit only on the lower-$\tau_E$ portion of the database can extrapolate to progressively higher-$\tau_E$ data. That construction was intended to mimic a low-performance $\rightarrow$ high-performance extrapolation task, and it showed that the lowest-order models ($N=1,2$) develop substantial far-tail error inflation, whereas models with $N \geq 3$ remain much more stable. A remaining limitation of that exercise, however, is that the $\tau_E$-ordered priors do not explicitly prevent the training set from containing data from machines that also appear in the held-out far tail. We therefore constructed a complementary extrapolation test in which the prior is imposed at the machine level rather than at the level of individual confinement-time slices. The question then becomes: how well would a given $N$-parameter scaling have predicted a more reactor-relevant machine using only information from machines ranked below it in a directed hierarchy?

Consistent with the previous subsubsection's fitting approach, for each model order $N \in \{1,\dots,9\}$ we retained the same preselected $N$-parameter variable set identified in the full-database model-selection step and, for each held-out target machine, refit only the coefficients of that fixed power-law form using data from the admissible source machines below it in the chosen hierarchy. We considered three such hierarchy schemes:
\begin{itemize}
    \item \texttt{A\_time\_order}: orders machines chronologically according to their earliest data record in DB5.2.3-STD5.
    \item \texttt{C1\_tau\_median}: orders machines by their machine-level median confinement time in the DB5.2.3-STD5.
    \item \texttt{C2\_param\_space}: orders machines by a composite reactorward score formed from the mean of robustly normalized machine-median values of $\{I_p, B_T, P_{l,th}, \kappa_a, \bar n_e, (1+\delta)\}$.
\end{itemize}

\begin{figure}[ht!]
    \begin{subfigure}[t]{0.99 \linewidth}
        \centering
        \includegraphics[width= 0.975\linewidth]{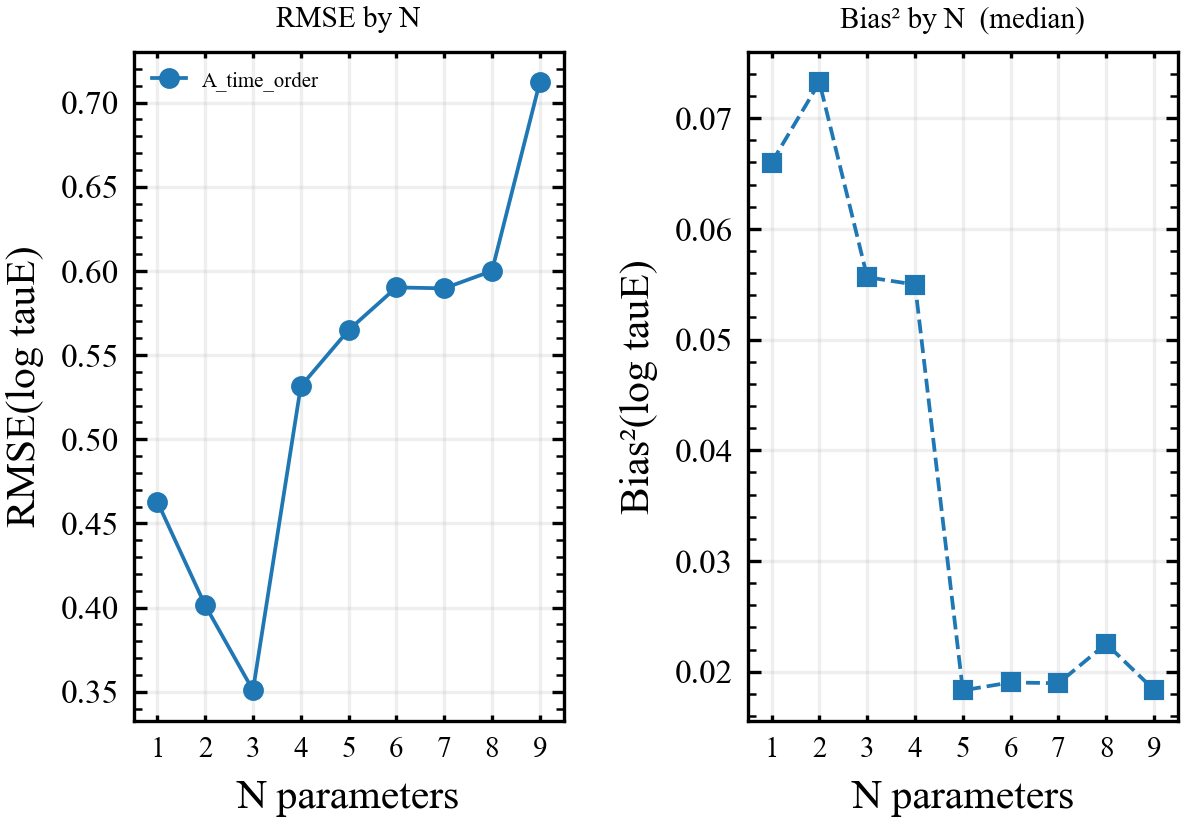}
        \label{subfig: A_time_order -- RMSE and Bias^2 vs. N-parameters }
    \end{subfigure}
    \begin{subfigure}[t]{0.99\linewidth}
        \centering
        \includegraphics[width= 0.97\linewidth]{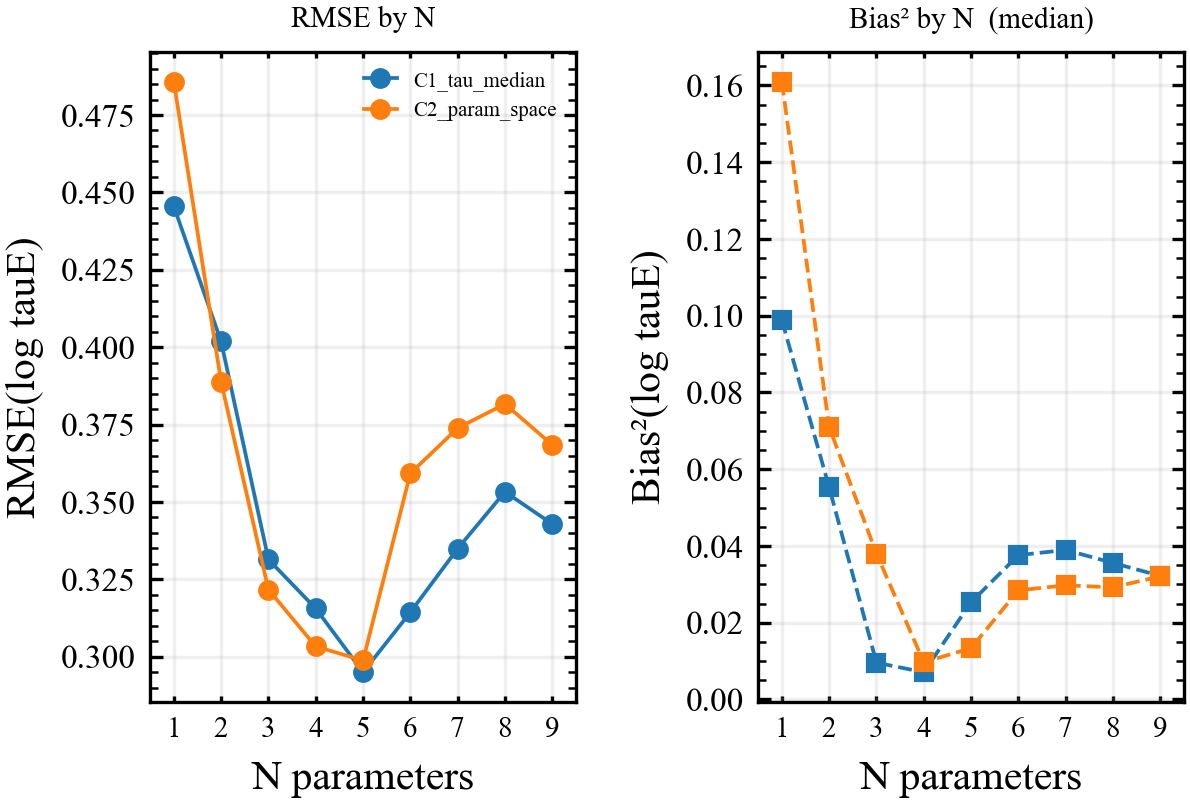}
         \label{subfig: C1 & C2 -- RMSE and Bias^2 vs. N-parameters}
    \end{subfigure}
    \captionsetup{width=\linewidth, font= footnotesize}
    \caption{Source-machine to held-out machine models' RMSE error and squared bias as a function of model complexity for several strategies, described in the main text.} 
    \label{fig: RMSE and Bias^2 vs. N-parameters}    
\end{figure}

For each hierarchy, the model trained on the allowed source machines was evaluated on the held-out subset of the next target machines. Extrapolative performance was quantified by the log-space $\mathrm{RMSE}$ and by a bias measure. The results are presented in figure \ref{fig: RMSE and Bias^2 vs. N-parameters} and discussed below. We note that the RMSEs in these machine-organized extrapolation tasks are expectedly larger than that exhibited by the final models in tables \ref{table: incremental_scalings_raw}--\ref{table: incremental_scalings_aug}.

The chronological hierarchy, \texttt{A\_time\_order}, is the most direct realization of the question: how well would a model of complexity $N$ have predicted the next tokamak in the historical development of the field using only its predecessors? Under this ordering, the extrapolative $\mathrm{RMSE}$ exhibits a particularly clear minimum at $N=3$. The error decreases sharply from the strongly underparameterized $N=1$ and $N=2$ cases to a global minimum at $N=3$, and then sharply increases as more parameters are added. The corresponding bias metric saturates to a minimum at $N=5$. Our assessment of this bias-variance is therefore that the variance penalty associated with moving beyond $N=3$ outweighs the bias minimum at $N=5$, so that the preferred truncation under the chronological criterion is $N=3$. This chronology test gives the clearest indication that increasing complexity beyond three parameters does not improve forward transferability, and in this configuration in fact degrades it in terms of RMSE. 

The machine-level confinement hierarchy, \texttt{C1\_tau\_median}, asks a slightly different question: how well can a model trained only on lower-confinement machines extrapolate to a machine with systematically higher characteristic confinement? This is in some ways similar to the last appendix subsection's question, except here the problem organizes $\tau_E$ by machine. Here, again, the lowest-order models are clearly inadequate, and the error improves rapidly as one moves into the low-to-moderate order model classes. In this hierarchy, the $\mathrm{RMSE}$ reaches a relatively shallow optimum at $N=5$, with $N=3$ and $N=4$ exhibiting marginally lower RMSE, then weakly degrades at increasing order. The bias metric under \texttt{C1\_tau\_median} saturates to an optimum around $N=3$ to $N=4$. In the bias-variance tradeoff, our interpretation is that the best compromise sits at the $N=4$ model class, with an eye toward the $N=3$. This indicates that when are ordered by a characteristic confinement level rather than by chronology, the fourth parameter appears to provide a small but non-negligible gain in balancing error inflation against systematic offset.

The composite reactorward hierarchy, \texttt{C2\_param\_space}, is intended to better reflect the multidimensional sense in which future reactor-relevant devices differ from the bulk of the historical database. Rather than ordering only by confinement level, it ranks machines by a composite score built from machine-median values of the principal engineering variables that emerge from the broader scaling analysis. In this case, the $\mathrm{RMSE}$ again improves rapidly with increasing complexity and saturates around $N=3$--$5$, beyond which the RMSE degrades. The bias metric reaches a shallow optimum at $N=3$, beyond which the bias degrades only weakly. That makes this hierarchy more aligned with the chronological ordering in its favoring of the $N=3$ model than with the characteristic confinement time ordering. 

The conclusions from each hierarchy are closely clustered. The chronological ordering most strongly favors $N=3$, because its error minimum is especially deep there and the rise in $\mathrm{RMSE}$ beyond $N=3$ is pronounced. The confinement-median hierarchy places the optimum closer to $N=4$, with both error and bias favoring that model order. The composite reactorward hierarchy again points primarily to $N=3$, while leaving $N=4$ clearly within the near-optimal range. When these views are taken together, the overall preferred truncation lies near $N=3$ to $N=4$.

This machine-organized variance-bias picture also helps contextualize the preceding appendix subsection's formal noninferiority result. There, the specific noninferiority test identified $N=3$ as the lowest complexity model noninferior to the $N=9$ reference in the far-tail metric, whereas $N=4$ did not satisfy that particular criterion. The present machine-forward analysis does not overturn that result, but it does show that $N=4$ sits consistently at or very near the optimal range once one examines the broader bias-variance tradeoff under machine-based extrapolation tasks with a strict cutoff. In that sense, the machine-hierarchy analysis redeems $N=4$ as a credible near-optimal model class even though it failed the narrower noninferiority test of the previous subsection. This is the reason $N=4$ is retained in the remainder of this work as a useful reference check alongside the principal $N=3$ model.

\subsection{Wall-discriminating coefficient variability with model complexity}
\label{appendix:  Wall-discriminating coefficient variability with model complexity}
\begin{figure}[ht!]
    \centering
    \captionsetup{width=\linewidth, font= footnotesize}
    \includegraphics[width=0.9\linewidth]{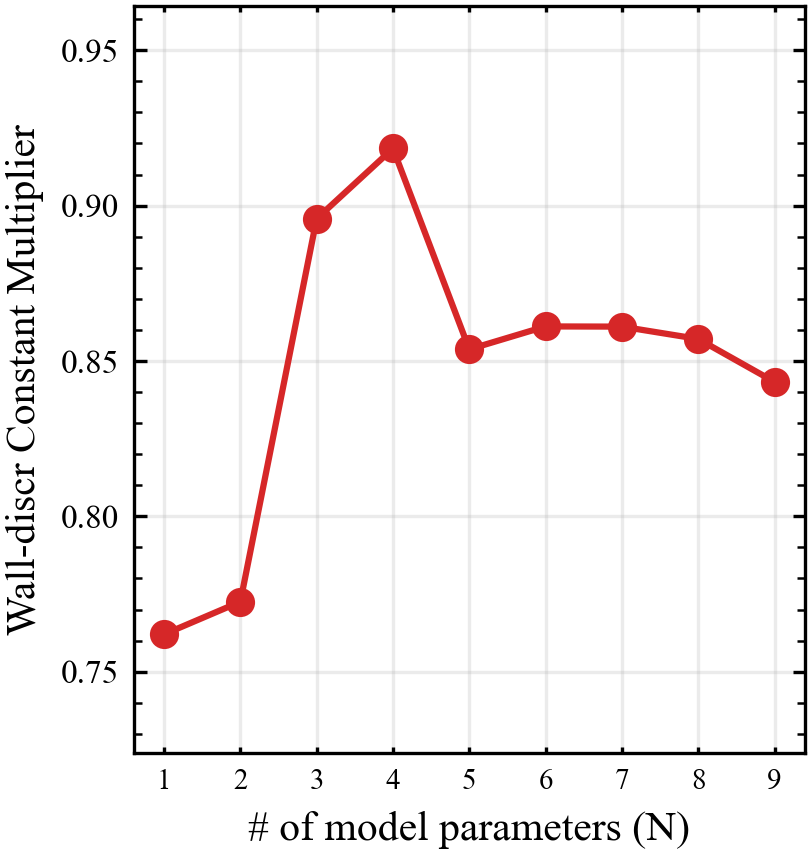}
    \caption{Wall-discriminating coefficient's variation with model complexity.}
    \label{fig: wall-constant variability up to N=9}
\end{figure}

It can be noted from table \ref{table: incremental_scalings_aug} that the wall-discriminating coefficient exhibits significant variation. Figure \ref{fig: wall-constant variability up to N=9} presents the convergence of the wall coefficient to approximately 0.85 as model complexity is increased up to the standard nine variables. This value can be interpreted as a measure of the confinement shortfall with high-Z walls with respect to low-Z walls. 

\subsection{$\langle nT \rangle \tau$ at constant $q_{95}$}
\begin{figure}[h]
    \centering
    \includegraphics[width=0.9\linewidth]{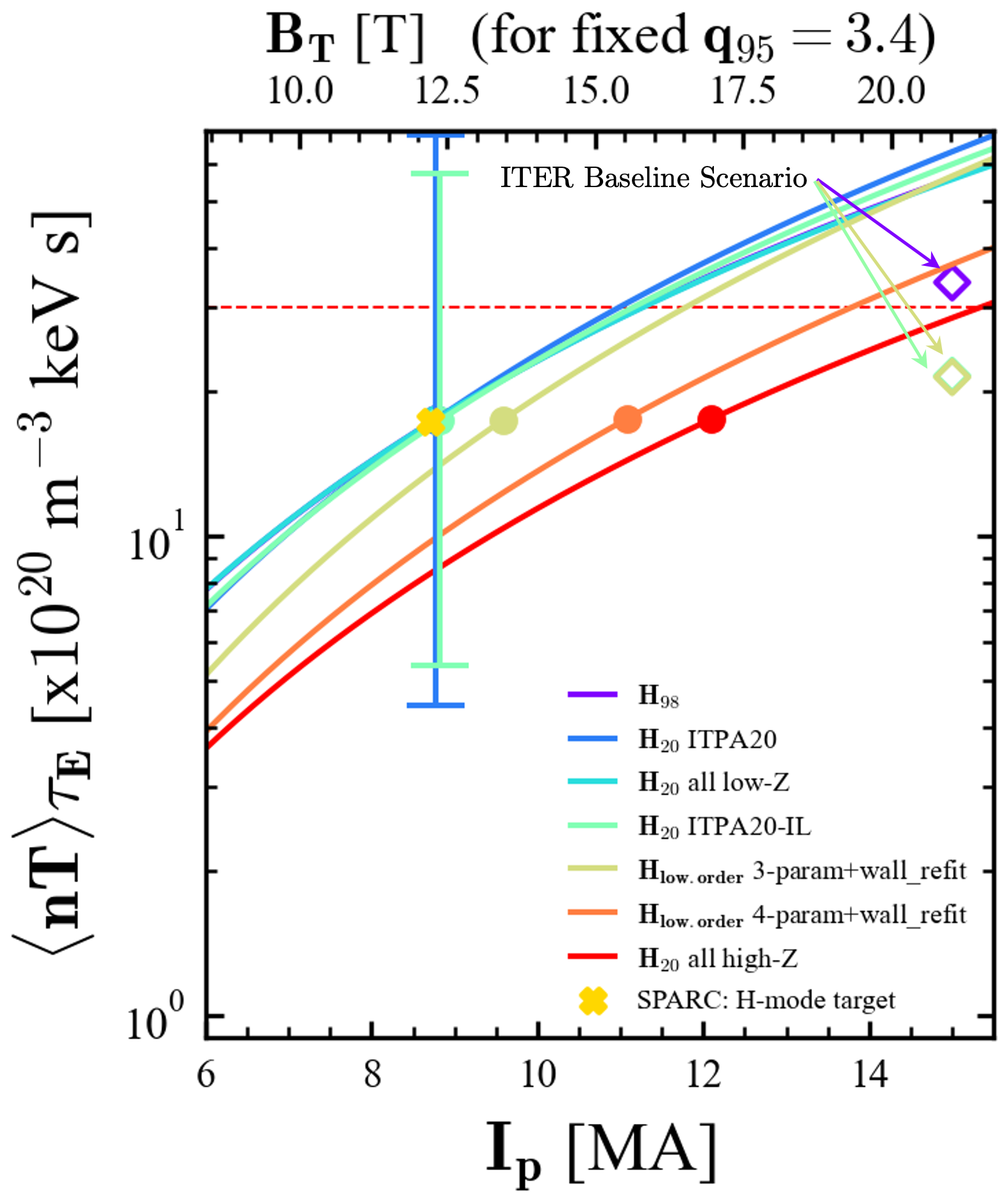}
    \caption{$\langle nT \rangle \tau_E$ for the SPARC H-mode scenario for an ensemble of confinement time scalings at fixed $q_{95}=3.4$, to be compared with figure \ref{fig: <nT>tau ensemble}. Note that the scaling color scheme is not identical with figure \ref{fig: <nT>tau ensemble}.}
    \label{fig: <nTtau> ensemble with fixed q=3.4}
\end{figure}

As noted at the end of section III, moving to higher $I_p$ runs into low-$q$ limits. Despite intrinsic macro-stability against tearing modes due to low $\beta$, instability can emerge if the edge $q$ is pushed too low. A reproduction of figure \ref{fig: <nT>tau ensemble} at constant $q=3.4$ is shown in figure \ref{fig: <nTtau> ensemble with fixed q=3.4}. 

\subsection{Fusion power scaling}
\label{appendix: Fusion power scaling}
\begin{figure}
    \centering
    \includegraphics[width=0.99\linewidth]{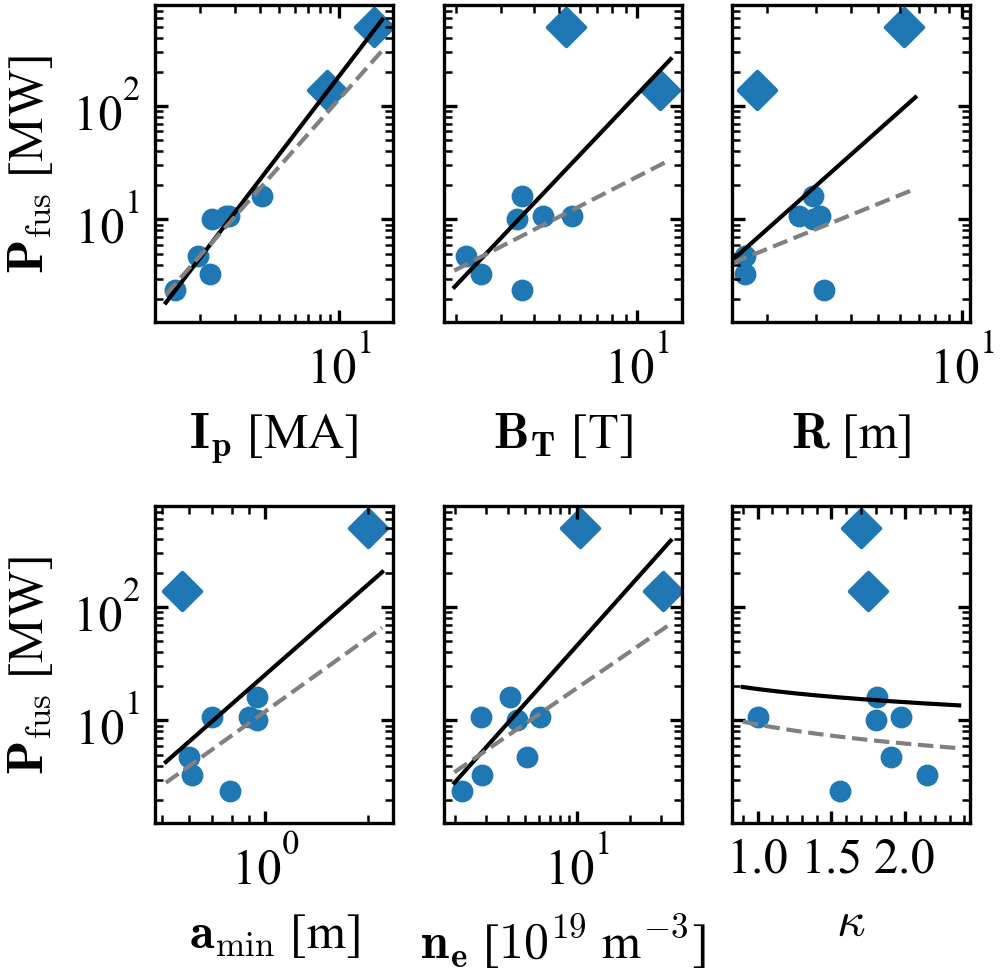}
    \caption{Single-variable power-law regressions with and without the machines currently under construction (produced from table \ref{table:  Pfus}) for variables of interest.}
    \label{fig: Pfus scaling for all variables}
\end{figure}

The other single-variable fits are provided as reference in table \ref{table: single_param_fits}. The underlying data and their power-law fits are presented in figure \ref{fig: Pfus scaling for all variables}.

\begin{table}[t]
\centering
\begin{tabular}{|c||cccc|}
\hline
\hline
Parameter & Exp (excl) & Exp (incl) & $R^2_{\log}$ (excl.) & $R^2_{\log}$ (incl) \\
\hline
$I_p$                  & 1.981  & 2.298  & 0.77 & 0.97 \\
$a_{\min}$             & 2.166  & 2.653  & 0.37 & 0.34 \\
$\bar{n}_e$            & 1.061  & 1.731  & 0.31 & 0.64 \\
$B_T$                  & 1.180  & 2.423  & 0.27 & 0.49 \\
$\kappa_a$             & -0.560 & -0.39 & 0.04 & 0.00 \\
\hline
\hline
\end{tabular}
\caption{Single-parameter power-law regression for fusion power. The "incl" columns include all points from table \ref{table:  Pfus}, while "excl." excludes the designs under construction (SPARC and ITER).}
\label{table: single_param_fits}
\end{table}

\subsection{Some expectations of fusion power's dependence on standard engineering parameters}
\label{appendix: Some expectations of fusion power's dependence on standard engineering parameters}

This appendix shows the wide-ranging expectations in terms of engineering parameters on how fusion power and $Q$ may be expected to scale. All expectations are derived for D-T and assumed to operate in the range of temperature in which the cross-section scales like $\propto T^2$. 

\subsubsection{Invoking the optimal low-order confinement scaling from this work ($N=3$ parameters)}
Approximating our optimal low-order scaling for the thermal energy confinement time as $\tau_{E,th} \propto I_p R_{geo}/P_{L,th}^{1/2}$; neglecting radiative losses; and assuming all of the thermal loss power comes from the auxiliary heating, fusion power takes the form,
\begin{align*}
    P_{fus} & \propto (nT)^2\,V \propto \frac{(P_{l,th} \,H\tau_E)^2}{V}  \propto   H^2\frac{I_p^{2} R^2}{P_{l,th}\,a^2} P_{l,th}^2  \propto H^2\,I_p^2 A^2 P_{l,th} \\ 
    P_{fus} &\propto  H^2 \,I_p^2 A^2 \frac{P_{aux}}{f_{aux}},
\end{align*}

where $f_{aux} = \frac{P_{aux}}{P_{l,th}}$ is the auxiliary heating fraction of the thermal loss power. With the same assumptions, $Q \propto H^2 \, I_p^2 A^2 /\ f_{aux}$. As $f_{aux} = P_{aux} / P_{l,th} \rightarrow 0$, $Q\rightarrow \infty$. 

A slightly different result is found when substituting the greenwald density, where, assuming $T$ is kept fixed for optimal DT cross-section, fusion power scales like $I_p^2/a^4$. 

\subsubsection{Resulting scaling from toroidal beta}
\label{subsubsection: Invoking toroidal beta}
This invokation finds that fusion power density scales quartically with the toroidal magnetic field and square of toroidal beta, $\beta_T = \frac{\langle p \rangle}{B_T^2/2\mu_0}$,

\begin{align*}
    P_{fus} \propto n^2T^2 \propto  p^2 \propto  \beta_T^2 B_T^4
\end{align*}

While it's true that $B_T$ and $\beta_T$ are technically independent quantities, in practice they are difficult to independently move at high performance, as pressure is typically the limiting quantity in high-field machines, less so traditional $\beta$ limits. 
 
\subsubsection{Resulting scaling from poloidal beta}
This invokation finds that fusion power density scales quartically with $I_p$ and quadratically with poloidal $\beta$, $\beta_p = \frac{\langle p \rangle}{I_p^2/2\mu_0}$. 

\begin{align*}
    P_{fus} \propto n^2T^2 \propto \beta_p^2 I_p^4. 
\end{align*}

As is the case with toroidal $\beta$, $\beta_p$'s dependence on $I_p^2$ makes $P_{fus}$'s parametric dependence difficult to interpret. Let us attempt to disambiguate this by substituting an expression for $\beta_p$. 

A common parameter optimized in fusion reactor designs in pursuit of minimizing the inductive current-drive is the bootstrap current, which scales like $j_{BS} \propto \frac{1}{B_p} \nabla p$.  The total bootstrap current over the cross-section is $I_{BS} \propto \frac{p}{B_p} \propto \frac{p}{I_p}$, since the poloidal field is generated by the plasma current. The bootstrap fraction is $f_{BS} = \frac{I_{BS}}{I_p} \propto \frac{p / I_p}{I_p} \propto \frac{p}{I_p^2} \propto \beta_p$. Substituting this back into fusion power density, we find the following expression,
\begin{align*}
    P_{fus} \propto f_{BS}^2 I_p^4. 
\end{align*}

This finding indicates that at fixed bootstrap fraction, fusion power density can be expected to scale quartically with the plasma current. This indicates a very strong dependence of fusion power on the plasma current. We can expect a generic, unfiltered tokamak database's tendency to decrease in bootstrap fraction at increasing plasma current due to the predominance of inductively-driven plasma current to ambiguate a single parameter fit in a fusion power scaling, perhaps collapsing to a discerned form closer to $\propto I_p^2$.

\subsection{Usage of auxiliary-heating dominance in self-consistent determination of $\langle nT \rangle$ in terms of $\tau_E$}

The $Q=11$ design-point is on the burn-supported branch of thermal equilibrium. Accessing it, an aspect of which is discussed in this work, generally requires traversing through the auxiliary-heated equilibrium branch. To this end, we crudely model the system as dominantly auxiliary-heated on the path toward $Q=11$. As such, values near and beyond $Q=11$ expect to suffer burn-related errors and more significantly deviate from the values presented herein.

\section{References}
\bibliographystyle{plain}
\small \bibliography{HighIp_is_King-Nov1925, every_reference}

\clearpage

\end{document}